\documentclass[journal]{IEEEtran}

\usepackage{cite}
\usepackage{amsmath}
\usepackage{amssymb}
\usepackage{amsfonts}
\usepackage{bm}
\usepackage{mathtools}
\usepackage{amsthm}
\usepackage{graphicx}
\usepackage{color}
\usepackage{textcomp}
\usepackage{grffile}

\usepackage{algorithm}
\usepackage{algpseudocode}

\makeatletter
\let\MYcaption\@makecaption
\makeatother
\usepackage{subcaption}
\makeatletter
\let\@makecaption\MYcaption
\makeatother

\newcommand{\minimize}{\operatornamewithlimits{\mathrm{minimize}}}
\newcommand{\argmin}{\operatornamewithlimits{\mathrm{arg\ min}}}
\newcommand{\subto}{\mathrm{subject\ to\ }}
\newcommand{\logit}{\mathrm{logit}}
\newcommand{\prox}{\mathrm{prox}}
\newcommand{\vect}{\mathrm{vec}}
\newcommand{\diag}{\mathrm{diag}}

\newcommand{\norm}[1]{\left\lVert#1\right\rVert}
\newcommand{\paren}[1]{\left(#1\right)}
\newcommand{\sqbra}[1]{\left[#1\right]}
\newcommand{\curbra}[1]{\left\{#1\right\}}

\usepackage{tikz}
\definecolor{myblue}{rgb}{0.184,0.392,0.620}
\definecolor{myorange}{rgb}{0.941,0.550,0.078}
\definecolor{mygreen}{rgb}{0,0.560,0}
\definecolor{myred}{rgb}{0.721,0.212,0.118}

\usetikzlibrary{shapes.geometric, arrows, positioning}
\tikzstyle{process} = [rectangle, minimum width=5cm, minimum height=1cm, text centered, text=white]
\tikzstyle{process_v} = [rectangle, minimum width=0.5cm, minimum height=2cm, text centered, text=black!70]
\tikzstyle{arrow} = [very thick,->,>=latex,draw=black!70]

\allowdisplaybreaks[1]

\begin{document}

\title{Deep Unfolding-Aided Parameter Tuning for Plug-and-Play-Based Video Snapshot Compressive Imaging}

\author{Takashi Matsuda, Ryo Hayakawa,~\IEEEmembership{Member,~IEEE,} and Youji Iiguni,~\IEEEmembership{Member,~IEEE}% <-this % stops a space
% \thanks{Manuscript received April 19, 2021; revised August 16, 2021.}%
\thanks{This work was supported by a research grant from The Mazda Foundation, a research grant from The Nakajima Foundation, and JSPS KAKENHI under Grant Number JP24K17277.}%
\thanks{Takashi Matsuda and Youji Iiguni are with Graduate School of Engineering Science, Osaka University, 1-3 Machikaneyama, Toyonaka, Osaka, 560-8531, Japan.}%
\thanks{Ryo Hayakawa is with Institute of Engineering, Tokyo University of Agriculture and Technology, 2-24-16 Naka-cho, Koganei, Tokyo, 184-8588, Japan (e-mail: hayakawa@go.tuat.ac.jp).}
}

\markboth{ACCEPTED TO IEEE ACCESS}%
{Deep Unfolding-Aided Parameter Tuning for Plug-and-Play-Based Video Snapshot Compressive Imaging}


\maketitle

\begin{abstract}
    Snapshot compressive imaging (SCI) captures high-dimensional data efficiently by compressing it into two-dimensional observations and reconstructing high-dimensional data from two-dimensional observations with various algorithms. 
    The plug-and-play (PnP) method is a promising approach for the video SCI reconstruction because it can leverage both observation models and denoising methods for videos. 
    Since the reconstruction accuracy significantly depends on the choice of noise level parameters, this paper proposes a deep unfolding-based method for tuning these parameters in PnP-based video SCI. 
    For the training of the parameters, we prepare training data from the densely annotated video segmentation dataset, reparametrize the noise level parameters, and apply the checkpointing technique to reduce the required memory. 
    Simulation results show that the trained noise level parameters via the proposed approach exhibit a non-monotonic pattern, which is different from the assumptions in the conventional convergence analyses of PnP-based algorithms. 
    These findings provide new insights into both the application of deep unfolding and the theoretical basis of PnP algorithms. 
\end{abstract}


%
%---------------
% Introduction
%---------------
\section{INTRODUCTION} \label{sec:intro}
\IEEEPARstart{S}{napshot} compressive imaging (SCI)~\cite{Yuan2021-eo} has gained attention as an approach to capture high-dimensional data. 
In SCI, high-dimensional data are modulated, compressed into two-dimensional data, and observed using a camera. 
We then reconstruct the original high-dimensional data using a reconstruction algorithm. 
As the data are compressed before observation, it is possible to acquire high-dimensional data with low memory and power consumption. 
SCI has been used in various fields, including video processing~\cite{Hitomi2011-hr,Reddy2011-bi,Llull2013-qn,Gao2014-wz}, hyperspectral imaging~\cite{Gehm2007-kd,Wagadarikar2008-ka,Wagadarikar2009-ph,Lin2014-pk,Yuan2015-hr,Pian2017-sa}, holography~\cite{Brady2009-zs,Zhang2018-we,Wang2017-op}, depth sensing~\cite{Llull2015-ve,Sun2017-bq}, and polarization imaging~\cite{Tsai2015-ad}. 

Video SCI is a technique used to obtain high frame rate videos by capturing information over multiple time frames. 
In video SCI, the unknown video of a scene is modulated by a mask and observed as a compressed image. 
The mask determines which pixels are observed in the compressed image and changes more rapidly than the frame rate of the camera. 
The high frame rate video is then reconstructed from the measurement image and the mask. 

The reconstruction methods for video SCI are mainly categorized into three types: mathematical optimization-based methods~\cite{Bioucas-Dias2007-vr,Yang2014-or,Yuan2016-hw,Liu2019-jg,Yang2020-yy}, machine learning-based methods~\cite{Qiao2020-xe,Cheng2021-wj,Cheng2023-yc,Miao2024-tj}, and hybrid methods~\cite{Yuan2020-wd,Yuan2022-nj,Wu2023-hb,Meng2023-xx,Liu2023-kj,Zhao2023-xe,Ma2023-nh,Shi2023-jw,Shi2024-uz}. 
In mathematical optimization-based methods, we first design an optimization problem and then construct an algorithm for the problem. 
We can incorporate various types of information on the measurement model and properties of videos directly in the design of the optimization problem. 
In contrast, in machine learning-based methods, we construct a machine learning model such as a deep neural network and train the model using data of the videos and the corresponding measurements. 
This approach has the potential to reconstruct unknown videos more accurately with lower computational complexity. 
For example, reversible SCI net~\cite{Cheng2021-wj} can achieve better reconstruction accuracy than a mathematical optimization-based method called decompress SCI (DeSCI)~\cite{Liu2019-jg}. 
However, learning-based methods require the training phase to train many parameters. 
Moreover, these methods often lack interpretability and flexibility for changes in the measurement model compared with mathematical optimization-based methods.

\IEEEpubidadjcol

To take advantage of both methods, hybrid approaches combining mathematical optimization and machine learning have been proposed for video SCI~\cite{Yuan2020-wd,Yuan2022-nj,Wu2023-hb,Meng2023-xx,Liu2023-kj,Zhao2023-xe,Ma2023-nh,Shi2023-jw,Shi2024-uz}. 
Most of these methods are based on the plug-and-play (PnP) approach~\cite{Venkatakrishnan2013-xj,Kamilov2023-tg}, where one of update equations of the optimization algorithm is replaced with a denoising method for natural images or videos. 
For example, on the basis of generalized alternating projection (GAP)~\cite{Liao2014-gy,Yuan2016-hw} and the alternating direction method of multipliers (ADMM)~\cite{Gabay1976-iu,Eckstein1992-ra,Combettes2011-id,Boyd2011-ci}, PnP-GAP and PnP-ADMM have been proposed for video SCI~\cite{Yuan2020-wd,Yuan2022-nj}. 
For the incorporated denoising method, various methods such as fast and flexible denoising network (FFDNet)~\cite{Zhang2018-zo} and fast deep video denoising net (FastDVDnet)~\cite{Tassano2020-os} have been utilized. 
These PnP methods have lower computational complexity than optimization-based DeSCI, whereas the learning-based high-quality denoising can be used in the algorithm. 

In PnP-based methods, we need to design noise level parameters, i.e., the strength of denoising at each iteration. 
The choice of the parameters significantly affects the final reconstruction accuracy. 
In conventional studies, however, the noise level parameters have been manually designed and have not been discussed deeply. 
To maximize the potential of PnP-based methods and achieve better reconstruction performance, we need to examine the effect of these parameters in more detail. 

In this paper, we propose a deep unfolding-based tuning method for the noise level parameters in the PnP approach. 
Deep unfolding~\cite{Gregor2010-xs,Balatsoukas-Stimming2019-yu,Monga2021-ph} leverages the similarity between the structure of iterative algorithms and neural networks to train appropriate values for internal algorithm parameters via automatic differentiation. 
Since the update equations of PnP-GAP and PnP-ADMM are differentiable with respect to the noise level parameters, deep unfolding can be used to train appropriate noise levels in principle. 
When we use deep learning-based denoising such as FastDVDnet, however, we require huge GPU memory to train the noise level parameters. 
To reduce the required memory, we use the checkpointing technique~\cite{Chen2016-mr,Gruslys2016-gz} in the proposed method. 
In computer simulations, we prepare the data for the training, train the noise level parameters, and evaluate the reconstruction performance using the conventional parameter settings. 
Simulation results show that the parameters trained via the proposed approach can achieve better reconstruction accuracy. 
Moreover, the trained parameters exhibit non-monotonic behavior during the iterations of the reconstruction algorithms. 
This implies that the monotonically decreasing noise levels assumed for the convergence proof~\cite{Chan2017-xv,Yuan2022-nj,Shi2023-jw,Shi2024-uz} are not necessarily optimal in terms of reconstruction accuracy. 
This interesting result is similar to the trained parameters of some other deep-unfolded algorithms~\cite{Ito2019-fm,Takabe2019-yc}. 
Our results would also provide insights into the research field of deep unfolding. 

The main contributions of this paper are summarized as follows.
\begin{itemize}
    \item We propose a parameter tuning method via deep unfolding for PnP-based reconstruction algorithms for video SCI. 
        By using the proposed approach, the noise level parameters can be trained with data and do not require complicated manual tuning. 
    \item For the training of PnP-based methods, we prepare the video data from the densely annotated video segmentation (DAVIS) dataset~\cite{Pont-Tuset2017-dr}, introduce the reparametrization for the noise level, and use the checkpointing approach in the training. 
        This approach for the training can be applied to PnP methods for other problems. 
    \item In the simulations, we show that the trained noise level parameters exhibit zigzag behavior, which is similar to those of other deep unfolding approaches~\cite{Ito2019-fm,Takabe2019-yc}. 
        This behavior is different from the usual assumption for the convergence guarantee~\cite{Chan2017-xv,Yuan2022-nj,Shi2023-jw,Shi2024-uz}; hence, our results provide motivation for the convergence analysis of PnP methods with non-decreasing noise level parameters. 
\end{itemize}

The rest of this paper is organized as follows. 
In Section~\ref{sec:system_model}, we describe the system model of video SCI. 
In Section~\ref{sec:conventional}, we review some conventional PnP-based methods for video SCI. 
In Section~\ref{sec:proposed}, we propose a parameter tuning method based on deep unfolding. 
In Section~\ref{sec:simulation}, we present the results of the computer simulations. 
Finally, we conclude the paper in Section~\ref{sec:conclusion}. 

In this paper, we use the following notation: 
$\mathbb{R}$ denotes the set of all real numbers, and $\mathbb{N}_{+}$ denotes the set of all positive integers. 
$(\cdot)^{\top}$ and $(\cdot)^{-1}$ represent the transpose and inverse of matrix, respectively. 
$\odot$ denotes the Hadamard product. 
For a vector $\bm{a} = [a_{1}, \dotsc, a_{N}]^{\top} \in \mathbb{R}^{N}$, $\norm{\bm{a}}_{2} = \sqrt{\sum_{n=1}^{N} a_{n}^{2}}$ denotes the $\ell_{2}$ norm of the vector, and $\diag(a_{1}, \dotsc, a_{N}) \in \mathbb{R}^{N \times N}$ denotes the diagonal matrix whose diagonal elements are $a_{1}, \dotsc, a_{N}$.
For a matrix $\bm{A} = \sqbra{\bm{a}_{1}\ \dotsb\ \bm{a}_{M}} \in \mathbb{R}^{N \times M}$, $\vect(\bm{A}) = \sqbra{\bm{a}_{1}^{\top} \ \dotsb \ \bm{a}_{M}^{\top}}^{\top} \in \mathbb{R}^{MN}$ denotes the vectorization of $\bm{A}$ obtained by stacking the columns of $\bm{A}$. 
For a function $f: \mathbb{R}^{N} \to \mathbb{R} \cup \{ +\infty \}$, the proximity operator of $f(\cdot)$ is defined as $\prox_{f}(\bm{s}) = \argmin_{\bm{x} \in \mathbb{R}^{N}} \curbra{f(\bm{x}) + \frac{1}{2} \norm{\bm{x} - \bm{s}}_{2}^{2}}$. 
\section{System Model} \label{sec:system_model}
\begin{figure}[t]
    \centering
    \includegraphics[width=1\columnwidth]{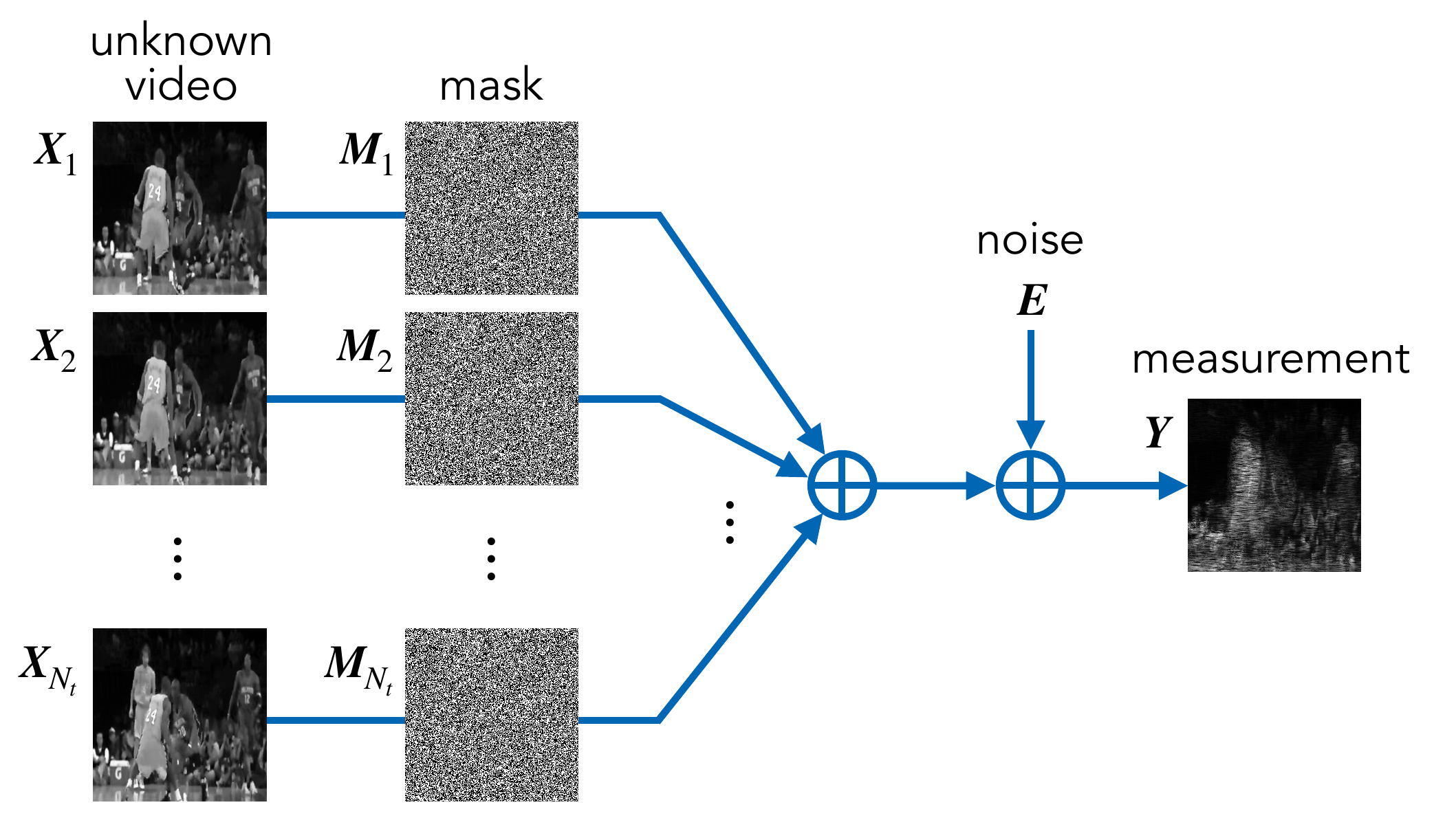}
    \caption{System model of video SCI.}
    \label{fig:model}
\end{figure}

In this section, we describe the system model of video SCI as shown in Fig.~\ref{fig:model}.
Let $\bm{X} \in \mathbb{R}^{N_{x} \times N_{y} \times N_{t}}$ represent an unknown grayscale video of $N_{t}$ frames, with an image size $N_{x} \times N_{y}$ corresponding to a scene. 
In SCI systems, the frames of the scene are modulated by a mask $\bm{M} \in \curbra{0, 1}^{N_{x} \times N_{y} \times N_{t}}$ and are observed as a two-dimensional compressed measurement 
\begin{align}
    \bm{Y} 
    &= 
    \sum_{n_{t}=1}^{N_{t}} \bm{X}_{n_{t}} \odot \bm{M}_{n_{t}} + \bm{E} \in \mathbb{R}^{N_{x} \times N_{y}}. \label{eq:model}
\end{align}
Here, $\bm{Y}$ is the measurement, $\bm{X}_{n_{t}} = \bm{X}(:, :, n_{t}) \in \mathbb{R}^{N_{x} \times N_{y}}$ denotes the $n_{t}$-th frame of the video, $\bm{M}_{n_{t}} = \bm{M}(:, :, n_{t}) \in \mathbb{R}^{N_{x} \times N_{y}}$ denotes the corresponding $n_{t}$-th mask, and $\bm{E} \in \mathbb{R}^{N_{x} \times N_{y}}$ denotes measurement noise. 
Since $\odot$ denotes the Hadamard product, one in the mask means that the corresponding pixel is observed, and zero means that the pixel is not observed. 

The measurement model in~\eqref{eq:model} can be transformed into the vectorized form as 
\begin{align}
    \bm{y} 
    &= 
    \bm{\Phi} \bm{x}^{\ast} + \bm{e}, \label{eq:model_vector}
\end{align}
where $\bm{y} = \vect(\bm{Y}) \in \mathbb{R}^{N_{x} N_{y}}$ and $\bm{e} = \vect(\bm{E}) \in \mathbb{R}^{N_{x} N_{y}}$ are the vectorized forms of the measurement image and noise, respectively. 
$\bm{\Phi} = [\bm{D}_{1}\ \dotsb\ \bm{D}_{N_{t}}] \in \mathbb{R}^{N_{x} N_{y} \times N_{x} N_{y} N_{t}}$ and $\bm{x}^{\ast} = \sqbra{\bm{x}_{1}^{\top}\ \dotsb\ \bm{x}_{N_{t}}^{\top}}^{\top} \in \mathbb{R}^{N_{x} N_{y} N_{t}}$ are the corresponding measurement matrix and the unknown vector, respectively, where $\bm{x}_{n_{t}} = \vect(\bm{X}_{n_{t}}) \in \mathbb{R}^{N_{x} N_{y}}$ represents the vectorization of the $n_{t}$-th frame and $\bm{D}_{n_{t}} = \diag (\vect(\bm{M}_{n_{t}})) \in \mathbb{R}^{N_{x} N_{y} \times N_{x} N_{y}}$ represents a diagonal matrix whose diagonal elements are the vectorized forms of $\bm{M}_{n_{t}}$ ($n_{t}=1,\dotsc,N_{t}$). 
The goal of video SCI is to reconstruct the unknown video $\bm{x}^{\ast}$ from the compressed measurement $\bm{y}$ and the information of the mask $\bm{\Phi}$. 
\section{Conventional Methods} \label{sec:conventional}
In TABLE~\ref{tab:related_work}, we summarize related work and recent advances in video SCI. 
In this section, we briefly review some conventional PnP-based methods that are closely related to the proposed approach. 
\begin{table*}[!t]
    \centering
    \caption{Summary of related work and recent advances in video SCI.}
    \label{tab:related_work}
    \renewcommand{\arraystretch}{1.5} % Increase line spacing
    \begin{tabular}{|c|c|p{11cm}|}
        \hline
        \textbf{Reference} & \textbf{Main Approach} & \textbf{Summary} \\ \hline
        J. Yang \textit{et al.}, 2014~\cite{Yang2014-or} & Model-based & Gaussian mixture model-based SCI for video reconstruction. \\ \hline
        X. Yuan \textit{et al.}, 2016~\cite{Yuan2016-hw} & Model-based & GAP algorithm for SCI. \\ \hline
        Y. Liu \textit{et al.}, 2019~\cite{Liu2019-jg} & Model-based & Rank minimization-based approach for SCI reconstruction. \\ \hline
        P. Yang \textit{et al.}, 2020~\cite{Yang2020-yy} & Model-based & SCI using Shearlet transform for improved sparsity representation. \\ \hline
        M. Qiao \textit{et al.}, 2020~\cite{Qiao2020-xe} & Machine Learning & Deep learning-based SCI reconstruction with convolutional architectures. \\ \hline
        Z. Cheng \textit{et al.}, 2021~\cite{Cheng2021-wj} & Machine Learning & Memory-efficient reversible SCI Net for large-scale video reconstruction. \\ \hline
        Z. Cheng \textit{et al.}, 2023~\cite{Cheng2023-yc} & Machine Learning & Recurrent neural network (RNN)-based SCI for efficient video reconstruction. \\ \hline
        Y.-C. Miao \textit{et al.}, 2024~\cite{Miao2024-tj} & Machine Learning & Domain factorization-based learning for SCI. \\ \hline
        X. Yuan \textit{et al.}, 2022~\cite{Yuan2022-nj} & Hybrid & PnP-GAP and PnP-ADMM algorithms for video SCI. \\ \hline
        Z. Wu \textit{et al.}, 2023~\cite{Wu2023-hb} & Hybrid & Adaptive parameter tuning for PnP methods to reduce the mismatch between training and test data. \\ \hline
        Z. Meng \textit{et al.}, 2023~\cite{Meng2023-xx} & Hybrid & Denoising network training via deep unfolding for SCI. \\ \hline
        X. Liu \textit{et al.}, 2023~\cite{Liu2023-kj} & Hybrid & Multi-stage fusion network for SCI with GAP framework. \\ \hline
        Y. Zhao \textit{et al.}, 2023~\cite{Zhao2023-xe} & Hybrid & Memory-efficient SCI reconstruction using deep equilibrium models. \\ \hline
        B. Shi \textit{et al.}, 2023~\cite{Shi2023-jw} & Hybrid & Provable denoising algorithms for SCI with convergence guarantees. \\ \hline
        B. Shi \textit{et al.}, 2024~\cite{Shi2024-uz} & Hybrid & Spatial-temporal video denoising methods for SCI with provable guarantees. \\ \hline
    \end{tabular}
\end{table*}
\subsection{PnP-ADMM} \label{subsec:admm}
The algorithm of PnP-ADMM is derived from the optimization problem
\begin{align}
    \minimize_{\bm{x} \in \mathbb{R}^{N_{x}N_{y}N_{t}}} 
    \curbra{
        \frac{1}{2} \norm{\bm{y} - \bm{\Phi} \bm{x}}_{2}^{2} + \lambda R (\bm{x})
    }. \label{eq:optimization_ADMM}
\end{align}
The first term in the problem in~\eqref{eq:optimization_ADMM} is a data fidelity term based on the measurement model in~\eqref{eq:model_vector}. 
The second term is a regularization term for videos and should be designed to return smaller values for clean videos. 
$\lambda$ ($> 0$) is a parameter that balances the data fidelity term and the regularization term.

To obtain the update equations of ADMM, the optimization problem is transformed to 
\begin{align}
    &\minimize_{\bm{x}, \bm{v} \in \mathbb{R}^{N_{x}N_{y}N_{t}}} 
    \curbra{
        \frac{1}{2} \norm{\bm{y} - \bm{\Phi} \bm{x}}_{2}^{2} + \lambda R (\bm{v})
    } \notag \\
    &\ \ \subto \  \bm{x} = \bm{v}. \label{eq:optimization_ADMM2}
\end{align}
The update equations of ADMM for the optimization problem~\eqref{eq:optimization_ADMM2} are given by
\begin{align}
    \bm{x}^{k+1} 
    &= 
    \argmin_{\bm{x} \in \mathbb{R}^{N_{x}N_{y}N_{t}}} 
    \curbra{\frac{1}{2} \norm{\bm{y} - \bm{\Phi} \bm{x}}_{2}^{2} + \frac{\rho}{2} \norm{\bm{x} - \bm{v}^{k} + \bm{u}^{k}}_{2}^{2}} \label{eq:ADMM_x} ,\\
    \bm{v}^{k+1} 
    &= 
    \argmin_{\bm{v} \in \mathbb{R}^{N_{x}N_{y}N_{t}}} 
    \curbra{\lambda R(\bm{v}) + \frac{\rho}{2} \norm{\bm{x}^{k+1} - \bm{v} + \bm{u}^{k}}_{2}^{2}} \label{eq:ADMM_v} ,\\
    \bm{u}^{k+1} 
    &= 
    \bm{u}^{k}+\bm{x}^{k+1}-\bm{v}^{k+1}, \label{eq:ADMM_u}
\end{align}
where $\rho$ ($> 0$) is the parameter.

The update equation of $\bm{x}$ in~\eqref{eq:ADMM_x} can be rewritten as
\begin{align}
    \bm{x}^{k+1} 
    = 
    \paren{\bm{\Phi}^{\top} \bm{\Phi} + \rho \bm{I}}^{-1}
    \paren{\bm{\Phi}^{\top} \bm{y} + \rho (\bm{v}^{k} - \bm{u}^{k})} \label{eq:ADMM_x2}.
\end{align}
Since the inverse matrix in~\eqref{eq:ADMM_x2} is a large matrix of size $N_{x}N_{y}N_{t} \times N_{x}N_{y}N_{t}$, direct computation of the inverse matrix $\paren{\bm{\Phi}^{\top} \bm{\Phi} + \rho \bm{I}}^{-1}$ is computationally expensive. 
However, the computation can be simplified by exploiting the structure of $\bm{\Phi}$. 
From the matrix inversion lemma, the update equation in~\eqref{eq:ADMM_x2} can be expressed as
\begin{align}
    \bm{x}^{k+1} 
    &= 
    \paren{\rho^{-1} \bm{I} - \rho^{-1} \bm{\Phi}^{\top} \paren{\bm{I} + \bm{\Phi} \rho^{-1} \bm{\Phi}^{\top}}^{-1} \bm{\Phi} \rho^{-1}} \notag \\
    &\hspace{5mm} 
    \times \paren{\bm{\Phi}^{\top} \bm{y} + \rho \paren{\bm{v}^{k} - \bm{u}^{k}}}. \label{eq:ADMM_x3}
\end{align}
From the definition of $\bm{\Phi}$, the matrix $\bm{\Phi} \bm{\Phi}^{\top}$ becomes diagonal and can be expressed as
\begin{align}
    \bm{\Phi}\bm{\Phi}^{\top} 
    &= 
    \diag \paren{\psi_{1}, \dots, \psi_{N_{x} N_{y}}}, \label{eq:PhiPhi}
\end{align}
where $\psi_{n} = \sum_{k=1}^{N_{t}} d_{k,n,n}$ and $d_{k, n, n}$ is the $(n, n)$ element of $\bm{D}_{k}$. 
Since \(\bm{\Phi}\bm{\Phi}^{\top}\) is a diagonal matrix, the inverse matrix in \eqref{eq:ADMM_x3} can be written as
\begin{align}
    \paren{\bm{I} + \bm{\Phi} \rho^{-1}\bm{\Phi}^{\top}}^{-1}
    = 
    \diag  \paren{\frac{\rho}{\rho + \psi_{1}}, \dots, \frac{\rho}{\rho + \psi_{N_{x}N_{y}}}}. 
\end{align}
Finally, the update in~\eqref{eq:ADMM_x3} can be rewritten as
\begin{align}
    \bm{x}^{k+1} 
    = 
    \bm{v}^{k} - \bm{u}^{k} 
    + \bm{\Phi}^{\top} \bm{\Psi} \bm{w}^{k+1}, \label{eq:ADMM_x4}
\end{align}
where 
\begin{align}
    \bm{\Psi} 
    = 
    \diag \paren{\frac{1}{\rho + \psi_{1}}, \dots, \frac{1}{\rho + \psi_{N_{x}N_{y}}}}
\end{align}
and 
\begin{align}
    \bm{w}^{k+1} 
    = 
    \bm{y} - \bm{\Phi} \paren{\bm{v}^{k} - \bm{u}^{k}}. \label{eq:ADMM_w}
\end{align}
By using~\eqref{eq:ADMM_x4} and~\eqref{eq:ADMM_w}, we can update $\bm{x}^{k}$ without computing the large-scale inverse matrix. 

The update of $\bm{v}^{k+1}$ in~\eqref{eq:ADMM_v} can be written as
\begin{align}
    \bm{v}^{k+1} 
    &= 
    \argmin_{\bm{v} \in \mathbb{R}^{N_{x}N_{y}N_{t}}} 
    \curbra{
        \lambda R (\bm{v}) + \frac{\rho}{2} \norm{ \bm{x}^{k+1} - \bm{v} + \bm{u}^{k} }^{2}_{2}
    } \label{eq:ADMM_v2} \\
    &= 
    \prox_{\frac{\lambda}{\rho} R} \paren{\bm{x}^{k+1}+\bm{u}^{k}}. \label{eq:ADMM_v3}
\end{align}
Since function $R(\cdot)$ is a regularizer for videos, it should return small values for clean videos. 
The right hand side of~\eqref{eq:ADMM_v2} indicates a clean video close to $\bm{x}^{k+1}+\bm{u}^{k}$, which can be interpreted as denoising for video $\bm{x}^{k+1}+\bm{u}^{k}$. 
From this viewpoint, the PnP approach replaces the update of $\bm{v}^{k}$ with denoising $\mathcal{D}_{\sigma_{k}}(\cdot)$, that is,
\begin{align}
    \bm{v}^{k+1} 
    = 
    \mathcal{D}_{\sigma_{k}} \paren{\bm{x}^{k+1} + \bm{u}^{k}}. \label{eq:ADMM_v4}
\end{align}
Here, $\sigma_{k} \in [0,1]$ is the noise level parameter used to determine the strength of denoising. 
As the denoising function $\mathcal{D}_{\sigma_{k}}(\cdot)$, we can use various denoising methods such as FFDNet~\cite{Zhang2018-zo} and FastDVDnet~\cite{Tassano2020-os}.
In~\cite{Yuan2022-nj}, FastDVDnet is used because it is trained for video denoising and can achieve high-quality denoising. 
The use of powerful denoising methods improves the reconstruction accuracy of the algorithm empirically. 
To obtain good reconstruction accuracy, however, the parameters $\sigma_{k} \in [0,1]$ ($k=0,1,\ldots$) need to be appropriately chosen. 

The algorithm of PnP-ADMM for video SCI is summarized in Algorithm~\ref{alg:PnP-ADMM}. 
\begin{algorithm}[tb]
    \caption{PnP-ADMM for video SCI}
    \label{alg:PnP-ADMM}
    \begin{algorithmic}[1]
        \Require $\bm{\Phi}$, $\bm{y}$, $\rho > 0$, $K \in \mathbb{N}_{+}$, $\{ \sigma_{k} \}_{k=0,1,\cdots,K-1} \in [0,1]$
        \Ensure $\bm{x}^{K}$
        \State Initialization: $\bm{v}^{0} = \bm{\Phi}^{\top} \bm{y}$, $\bm{u}^{0} = \bm{0}$
        \State Compute $\psi_{n}$ ($n=1,\dotsc,N_{x}N_{y}$) from~\eqref{eq:PhiPhi}
        \State $\bm{\Psi} = \diag \paren{\frac{1}{\rho + \psi_{1}}, \dots, \frac{1}{\rho + \psi_{N_{x}N_{y}}}}$
        \For{$k=0,1,\cdots,K-1$}
            \State $\bm{w}^{k+1} = \bm{y} - \bm{\Phi} \paren{\bm{v}^{k} - \bm{u}^{k}}$
            \State $\bm{x}^{k+1} = \bm{v}^{k} - \bm{u}^{k} + \bm{\Phi}^{\top} \bm{\Psi} \bm{w}^{k+1}$
            \State $\bm{v}^{k+1} = \mathcal{D}_{\sigma_{k}}(\bm{x}^{k+1}+\bm{u}^{k})$
            \State $\bm{u}^{k+1} = \bm{u}^{k}+\bm{x}^{k+1}-\bm{v}^{k+1}$
        \EndFor
    \end{algorithmic}
\end{algorithm}
\subsection{PnP-GAP} \label{subsec:gap}
PnP-GAP is derived from the optimization problem
\begin{align}
    &\minimize_{\bm{x}, \bm{v} \in \mathbb{R}^{N_{x}N_{y}N_{t}}}
    \curbra{
        \frac{1}{2} \norm{ \bm{x} - \bm{v} }_{2}^{2} + \lambda R (\bm{v})
    } \notag \\
    &\ \ \subto \ \bm{y} = \bm{\Phi} \bm{x}. \label{eq:optimization_GAP}
\end{align}
The constraint $\bm{y} = \bm{\Phi} \bm{x}$ restricts the variable $\bm{x}$ by using the measurement model in~\eqref{eq:model_vector}. 
The variable $\bm{v}$ is optimized by minimizing the sum of the distance from $\bm{x}$ and the regularizer $R(\cdot)$ for videos. 
In PnP-GAP, we alternately update $\bm{x}$ and $\bm{v}$ in accordance with~\eqref{eq:optimization_GAP}. 
The update equations of PnP-GAP are thus given by
\begin{align}
    \bm{x}^{k+1} 
    &= 
    \argmin_{\bm{x} \in \mathbb{R}^{N_{x}N_{y}N_{t}}} 
    \curbra{ \frac{1}{2} \norm{\bm{x} - \bm{v}^{k}}_{2}^{2}} \notag \\
    &\hspace{7mm} \subto \ \bm{y} = \bm{\Phi}\bm{x} \label{eq:GAP_x}, \\
    \bm{v}^{k+1} 
    &= 
    \argmin_{\bm{v} \in \mathbb{R}^{N_{x}N_{y}N_{t}}}
    \curbra{ 
        \frac{1}{2} \norm{\bm{x}^{k+1} - \bm{v}}_{2}^{2} 
        + \lambda R (\bm{v}) 
    }. \label{eq:GAP_v}
\end{align}

The update of $\bm{x}^{k}$ in~\eqref{eq:GAP_x} can be written as 
\begin{align}
    \bm{x}^{k+1} 
    = 
    \bm{v}^{k} 
    + \bm{\Phi}^{\top} \paren{\bm{\Phi} \bm{\Phi}^{\top}}^{-1} 
    \paren{ \bm{y} - \bm{\Phi} \bm{v}^{k} }.
    \label{eq:GAP_x2}
\end{align}
Since $\bm{\Phi}\bm{\Phi}^{\top}$ is a diagonal matrix as discussed in~\eqref{eq:PhiPhi}, the update in~\eqref{eq:GAP_x2} can be further rewritten as 
\begin{align}
    \bm{x}^{k+1} 
    = 
    \bm{v}^{k} 
    + 
    \bm{\Phi}^{\top}
    \tilde{\bm{\Psi}} \bm{w}^{k+1}, \label{eq:GAP_x3}
\end{align}
where 
\begin{align}
    \tilde{\bm{\Psi}} 
    = 
    \diag \paren{ \frac{1}{\psi_{1}}, \dotsc, \frac{1}{\psi_{N_{x}N_{y}}}}
\end{align}
and 
\begin{align}
    \bm{w}^{k+1} 
    = 
    \bm{y} - \bm{\Phi} \bm{v}^{k}. \label{eq:GAP_w}
\end{align}
Hence, the calculation of the inverse matrix in~\eqref{eq:GAP_x2} can be replaced with element-wise calculations to efficiently compute $\bm{x}^{t+1}$. 

As in PnP-ADMM, the update of $\bm{v}^{k}$ in~\eqref{eq:GAP_v} is replaced with a denoising function $\mathcal{D}_{\sigma_{k}}(\cdot)$, i.e.,
\begin{align}
    \bm{v}^{k+1} 
    = 
    \mathcal{D}_{\sigma_k}(\bm{x}^{k+1}). \label{eq:GAP_v2}
\end{align}
Note again that it is necessary to properly set the parameter $\sigma_k \in [0,1]$ at each iteration. 

In~\cite{Liao2014-gy,Yuan2016-hw}, the convergence of GAP is accelerated by replacing $\bm{y}$ in~\eqref{eq:GAP_w} with 
\begin{align}
    \bm{y}^{k+1} 
    = 
    \bm{y}^{k} + \paren{\bm{y} - \bm{\Phi}\bm{v}^{k}} \label{eq:GAP_y}
\end{align}
at each iteration. 
The accelerated version of PnP-GAP is summarized as 
\begin{align}
    \bm{y}^{k+1} 
    &= 
    \bm{y}^{k} + \paren{\bm{y} - \bm{\Phi} \bm{v}^{k}}, \\
    \bm{w}^{k+1} 
    &= 
    \bm{y}^{k+1} - \bm{\Phi} \bm{v}^{k}, \\
    \bm{x}^{k+1} 
    &= 
    \bm{v}^{k} 
    + 
    \bm{\Phi}^{\top} \tilde{\bm{\Psi}} \bm{w}^{k+1}, \\
    \bm{v}^{k+1} 
    &= 
    \mathcal{D}_{\sigma_{k}}\paren{\bm{x}^{k+1}}.
\end{align}
The detailed algorithm is shown in Algorithm~\ref{alg:PnP-GAP}. 
\begin{algorithm}[tb]
    \caption{PnP-GAP (accelerated version) for video SCI}
    \label{alg:PnP-GAP}
    \begin{algorithmic}[1]
        \Require $\bm{\Phi}$, $\bm{y}$, $K \in \mathbb{N}_{+}$, $\{ \sigma_{k} \}_{k=0,1,\cdots,K-1} \in [0,1]$
        \Ensure $\bm{x}^{K}$
        \State Initialization: $\bm{v}^{0} = \bm{\Phi}^{\top} \bm{y}$, $\bm{y}^{0} = \bm{0}$
        \State Compute $\psi_{n}$ ($n=1,\dotsc,N_{x}N_{y}$) from~\eqref{eq:PhiPhi}
        \State $\tilde{\bm{\Psi}} = \diag \paren{\frac{1}{\psi_{1}}, \dots, \frac{1}{\psi_{N_{x}N_{y}}}}$
        \For{$k=0,1,\cdots,K-1$}
            \State $\bm{y}^{k+1} = \bm{y}^{k} + \paren{\bm{y} - \bm{\Phi} \bm{v}^{k}}$
            \State $\bm{w}^{k+1} = \bm{y}^{k+1} - \bm{\Phi} \bm{v}^{k}$
            \State $\bm{x}^{k+1} = \bm{v}^{k} + \bm{\Phi}^{\top} \tilde{\bm{\Psi}} \bm{w}^{k+1}$
            \State $\bm{v}^{k+1} = \mathcal{D}_{\sigma_{k}} \paren{\bm{x}^{k+1}}$
        \EndFor
    \end{algorithmic}
\end{algorithm}
\subsection{Noise Level Paremeters in PnP Methods} \label{subsec:param}
In PnP-ADMM and PnP-GAP, we have the parameters $\curbra{\sigma_{k}}_{k=0, \dotsc, K-1}$ to control the strength of denoising at each iteration. 
The setting of these parameters significantly affects the accuracy of the video reconstruction. 
Although FFDNet and FastDVDnet can use a noise level map of the same size as the input image, the noise level parameters in PnP methods are usually treated as scalar values, which means that each element of the noise level map has the same value among all pixels. 
In the previous study~\cite{Yuan2022-nj}, the following three settings are considered for 60 iterations: 
\begin{enumerate}
    \item \textbf{step}: $\sigma_{0} = \dotsb = \sigma_{19} = 50/255$, $\sigma_{20} = \dotsb = \sigma_{39} = 25/255$, $\sigma_{40} = \dotsb = \sigma_{59} = 12/255$
    \item \textbf{exponential}: $\sigma_{0} = 50/255, \sigma_{k} = 0.97  \sigma_{k-1}$ ($k = 1, \dotsc, K-1$)
    \item \textbf{constant}: $\sigma_{k} = 12/255$ ($k = 0,1,\dotsc,K-1$)
\end{enumerate}
Particularly in cases without preprocessing, it has been found in~\cite{Yuan2022-nj} that the first setting (step) achieves the best reconstruction accuracy in simulations. 
However, the setting is not necessarily optimal, and we need to tune the parameters $\curbra{\sigma_{k}}_{k=0, \dotsc, K-1}$ manually to obtain better reconstruction accuracy. 

\subsection{Other Related Work} \label{subsec:related}
Several PnP-based approaches have also been proposed for video SCI. 
In~\cite{Wu2023-hb}, the parameter in the denoising network is updated adaptively in the reconstruction to mitigate the difference between the training data and reconstructed videos. 
Since the noise level parameters are still treated as hyperparameters, the focused point is different from our approach for training the noise level parameters. 
We might be able to combine our proposed approach with the method in~\cite{Wu2023-hb}. 
In~\cite{Meng2023-xx}, a denoising network with the structure of auto-encoder, DnCNN, U-net, or ResNet is considered. 
The parameters in the network are then trained via deep unfolding. 
In this approach, however, we have to train many parameters compared with our approach in Section~\ref{sec:proposed}. 
Moreover, we need to design a highly complicated denoising network to use the information between multiple frames in the video. 
A GAP-based network~\cite{Liu2023-kj} achieves better reconstruction accuracy than other methods, whereas large memory consumption might be a problem for a high compression rate. 
In~\cite{Zhao2023-xe}, a reconstruction method based on deep equilibrium models has been proposed. 
This memory efficient approach is one solution for the training of the PnP-based method. 
The trained parameters in this approach remains unchanged over the algorithm iterations, whereas our focus is on the design of the parameter values at each iteration. 
\begin{figure*}[t]
    \begin{minipage}[t]{0.3\hsize}
        \centering
        \scalebox{0.8}{
            \begin{tikzpicture}[node distance=1.6cm]
                % Outer rectangle
                \draw [fill=gray!20, draw=none, rounded corners] (-3.8,1) rectangle (2.8,-6.3);
                % Nodes
                \node (procA) [process, fill=myblue!90] {$\bm{w}^{k+1} = \bm{y} - \bm{\Phi} \paren{\bm{v}^{k} - \bm{u}^{k}}$};
                \node (procB) [process, below of=procA, fill=myorange!90] {$\bm{x}^{k+1} = \bm{v}^{k} - \bm{u}^{k} + \bm{\Phi}^{\top} \bm{\Psi} \bm{w}^{k+1}$};
                \node (procC) [process, below of=procB, fill=mygreen!90] {$\bm{v}^{k+1} = \mathcal{D}_{\sigma_{k}}(\bm{x}^{k+1}+\bm{u}^{k})$};
                \node (procD) [process, below of=procC, fill=myred!90] {$\bm{u}^{k+1} = \bm{u}^{k}+\bm{x}^{k+1}-\bm{v}^{k+1}$};
                
                % Arrows
                \draw [arrow] (procA) -- (procB)  node[midway, right, xshift=0.5mm] {$\bm{w}^{k+1}$};
                \draw [arrow] (procB) -- (procC)  node[midway, right, xshift=0.5mm] {$\bm{x}^{k+1}$};
                \draw [arrow] (procC) -- (procD)  node[midway, right, xshift=0.5mm] {$\bm{v}^{k+1}$};
                \draw [arrow] (procD.south) -- ++(0,-0.5) node[midway, right, xshift=0.5mm] {$\bm{u}^{k+1}$} -- ++(-3.5,0) -- ++(0,5.8) -- (procA.west);
            \end{tikzpicture}
        }
        \subcaption{Signal flow of PnP-ADMM.}
        \label{subfig:ADMM}
    \end{minipage}
    \begin{minipage}[t]{0.7\hsize}
        \centering
        \scalebox{0.8}{
            \begin{tikzpicture}[node distance=1.4cm]
                % Outer rectangle
                \draw [fill=gray!20, draw=none, rounded corners] (-0.6,3.15) rectangle (14.3,-4.15);

                % Nodes
                \node (procA) [process_v, fill=myblue!90] {};
                \node (procB) [process_v, right of=procA, fill=myorange!90] {};
                \node (procC) [process_v, right of=procB, fill=mygreen!90] {};
                \node (procD) [process_v, right of=procC, fill=myred!90] {};
                \node (proc_space) [process_v, right of=procD, fill=gray!20] {$\cdots$};
                \node (procC2) [process_v, right of=proc_space, fill=mygreen!90] {};
                \node (procD2) [process_v, right of=procC2, fill=myred!90] {};
                \node (procA2) [process_v, right of=procD2, fill=myblue!90] {};
                \node (procB2) [process_v, right of=procA2, fill=myorange!90] {};
                \node (loss) [right=8mm of procB2] {$\mathcal{L}\paren{\bm{x}^{K}, \bm{x}^{\ast}}$};

                \node (sigma0) at (procC.south) [circle, fill=purple, inner sep=4pt, yshift=2.7mm] {};
                \node[below right=1mm and 0.5mm of sigma0, text=purple] {$\sigma_{0}$};
                \node (sigmaK-1) at (procC2.south) [circle, fill=purple, inner sep=4pt, yshift=2.7mm] {};
                \node[below right=1mm and 0.5mm of sigmaK-1, text=purple] {$\sigma_{K-2}$};

                % Arrows
                \draw [arrow] (procA) -- (procB) node[midway, above, yshift=0.5mm] {$\bm{w}^{1}$};
                \draw [arrow] (procB) -- (procC) node[midway, above, yshift=0.5mm] {$\bm{x}^{1}$};
                \draw [arrow] (procC) -- (procD) node[midway, above, yshift=0.5mm] {$\bm{v}^{1}$};
                \draw [arrow] (procD) -- (proc_space) node[midway, above, yshift=0.5mm] {$\bm{u}^{1}$};
                \draw [arrow] (proc_space) -- (procC2) node[midway, above, yshift=0.5mm, xshift=-1mm] {$\bm{x}^{K-1}$};
                \draw [arrow] (procC2) -- (procD2) node[midway, above, yshift=0.5mm] {$\bm{v}^{K-1}$};
                \draw [arrow] (procD2) -- (procA2) node[midway, above, yshift=0.5mm] {$\bm{u}^{K-1}$};
                \draw [arrow] (procA2) -- (procB2) node[midway, above, yshift=0.5mm] {$\bm{w}^{K}$};
                \draw [arrow] (procB2) -- (loss) node[midway, above, yshift=0.5mm, xshift=1mm] {$\bm{x}^{K}$};
                \draw [arrow, dashed] (loss) -- ++(0,-2) -- ++(-10.42,0) -- (sigma0.south);
                \draw [arrow, dashed] (loss) -- ++(0,-2) -- ++(-6.22,0) node[midway, below] {update by backpropagation} -- (sigmaK-1.south);
            \end{tikzpicture}
        }
        \subcaption{Unfolded signal flow of PnP-ADMM.}
        \label{subfig:unfolded_ADMM}
    \end{minipage}
    \caption{Signal flow of PnP-ADMM and its unfolded version.}
    \label{fig:ADMM_unfolding}
\end{figure*}
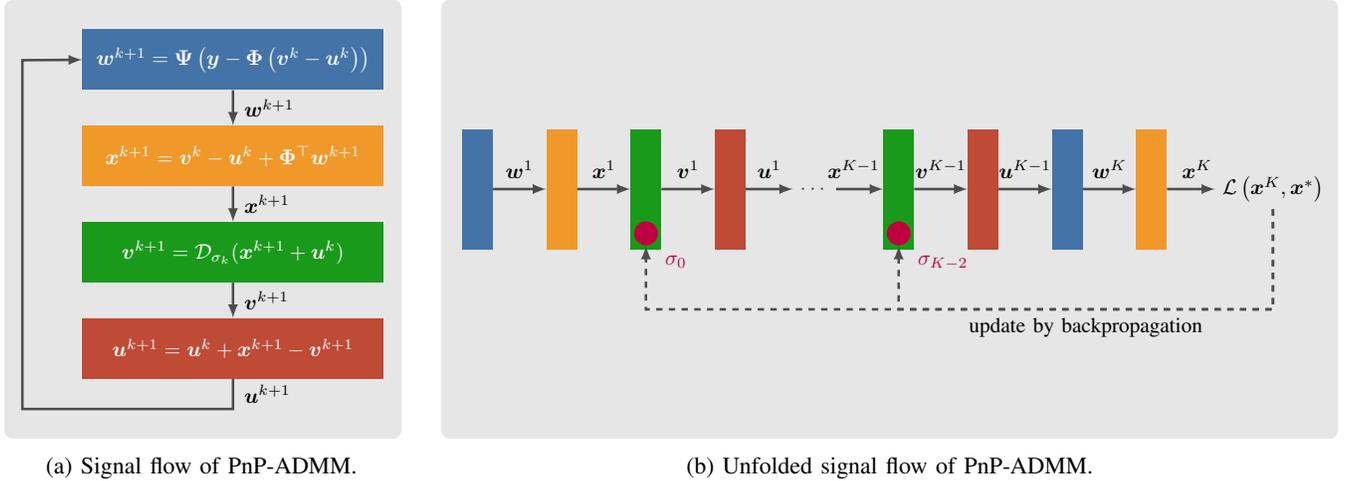
\section{Proposed Parameter Tuning via Deep Unfolding} \label{sec:proposed}
In~\cite{Yuan2022-nj}, the noise level parameter $\sigma_{k}$ for the denoising method is designed manually. 
To obtain better parameter choice, we propose a data-driven approach based on deep unfolding~\cite{Gregor2010-xs,Balatsoukas-Stimming2019-yu,Monga2021-ph}. 

\subsection{Proposed Network via Deep Unfolding} \label{subsec:deep_unfolding}
Deep unfolding is a method to train the parameter values in iterative algorithms by leveraging deep learning techniques such as backpropagation and stochastic gradient descent. 
Pioneered by the work of learned iterative shrinkage thresholding algorithm (LISTA)~\cite{Gregor2010-xs}, this approach has been applied to various problems such as sparse signal recovery, image restoration, and wireless communication~\cite{Yang2016-di,Adler2018-xv,Ito2019-fm,Takabe2019-yc,Wadayama2019-wz,Li2020-wi,Nagahama2022-yl}. 
Deep unfolding can be regarded as an integration of mathematical optimization and machine learning from a different perspective of PnP. 
Compared to the conventional approach of manual setting, deep unfolding can automatically train the parameters from the data. 
Moreover, the number of trainable parameters in deep unfolding is smaller than that in usual deep neural networks, and the interpretability of the result is clearer. 

In this study, we use the deep unfolding approach to tune the parameters $\curbra{\sigma_{k}}_{k=0,\ldots,K-1}$ in PnP-ADMM and PnP-GAP.
Reasonable parameter values for iterative algorithms should minimize a loss function $\mathcal{L}\paren{\hat{\bm{x}},\bm{x}^{\ast}}$, which represents the difference between the estimate $\hat{\bm{x}}$ and its true value $\bm{x}^{\ast}$. 
For the deep unfolding-based parameter tuning, we prepare the data $\paren{\bm{y}_{(1)}, \bm{x}_{(1)}^{\ast}}, \dotsc, \paren{\bm{y}_{(J)}, \bm{x}_{(J)}^{\ast}}$ for $\paren{\bm{y}, \bm{x}^{\ast}}$ and train the parameters $\curbra{\sigma_{k}}_{k=0,\dotsc,K-1}$ by minimizing the loss function, where $J$ is the number of data. 
Specifically, we consider the optimization problem 
\begin{align}
    \minimize_{\curbra{\sigma_{k}}_{k=0,\dotsc,K-1}} 
    \curbra{
        \sum_{j=1}^{J} \mathcal{L}\paren{\hat{\bm{x}}_{(j)},\bm{x}_{(j)}^{\ast}}
    }  \label{eq:optimization_deep_unfolding}
\end{align}
and tune the parameter based on the optimization problem in~\eqref{eq:optimization_deep_unfolding}. 

To train the parameters $\curbra{\sigma_{k}}_{k=0,\dotsc,K-1}$, we need to compute the gradient of the loss function with respect to the parameters. 
The idea of deep unfolding is to regard the iterative algorithm as a deep neural network and compute the gradient by backpropagation. 
For example, the signal flow of PnP-ADMM and its unfolded version are shown in Fig.~\ref{fig:ADMM_unfolding}. 
Note that some signal flows are omitted for simplicity. 
If the denoiser $\mathcal{D}_{\sigma_{k}}(\cdot)$ is differentiable with respect to $\sigma_{k}$, the gradient of the loss function can also be computed by backpropagation, because each iteration of the algorithm can be differentiable. 
\subsection{Details of Training} \label{subsec:training}
\subsubsection{Dataset} \label{subsubsec:dataset}
As the training dataset, we use a modified version of the DAVIS dataset~\cite{Pont-Tuset2017-dr}. 
The DAVIS dataset was converted into grayscale videos and downsampled so that the height is 256 pixels. 
The width was then cut to 256 pixels by taking the central part. 
For all videos, $N_{t} = 8$ frames of the video were compressed into a single image. 
The maximum pixel value of the images is normalized to one. 
For each observed image, we added additive Gaussian noise with zero mean and standard deviation of $0.01$. 
For the training data, we use 671 observed images (corresponding to 5,368 unknown frames) in the modified DAVIS dataset. 
As the test dataset, we use six grayscale videos (\texttt{kobe}, \texttt{traffic}, \texttt{runner}, \texttt{drop}, \texttt{crash}, \texttt{aerial})~\cite{Liu2019-jg}, which are typically used for the performance evaluation of video SCI algorithms. 
For the mask $\bm{M}$, we use the same mask as that in the dataset of the six grayscale videos. 
\subsubsection{Reparametrization} \label{subsubsec:reparametrization}
Since the noise level parameter $\sigma_{k}$ is in the range $[0,1]$, it is necessary to ensure that $\sigma_{k}$ remains within this range. 
Hence, we use the output of the sigmoid function as $\sigma_{k}$, i.e.,  
\begin{align}
    \sigma_{k} 
    &= 
    \frac{1}{1 + e^{-p_k}}, \label{eq:sigmoid}
\end{align}
where $p_{k}$ is the actual trained parameter. 
Note that $\sigma_{k}$ in~\eqref{eq:sigmoid} is within the range $(0, 1)$. 
For comparison, the number of iterations in PnP-ADMM and PnP-GAP is set to 60. 
The initial value of $p_{k}$ in the training is set such that the noise level parameter $\sigma_{k}$ follows the first setting in Section~\ref{subsec:param}. 
This means 
\begin{align}
    p_{0} &= \dotsb = p_{19} = \logit(50/255), \\
    p_{20} &= \dotsb = p_{39} = \logit(25/255), \\
    p_{40} &= \dotsb = p_{59} = \logit(12/255),
\end{align}
where the logit function $\mathrm{logit}(\sigma) = \log\paren{\dfrac{\sigma}{1-\sigma}}$ is the inverse function of the sigmoid function. 
\subsubsection{Training} \label{subsubsec:training}
As the optimizer, we use the Adam optimizer~\cite{Kingma2014-fy} with a learning rate of $0.01$. 
To reduce computational complexity, we do not use incremental training~\cite{Ito2019-fm}, which is often used to prevent the vanishing gradient problem in deep unfolding. 
The number of iterations is set to $K = 60$, the parameter of ADMM is $\rho = 0.01$, the minibatch size is $5$, and the number of epochs is $10$. 
The loss function $\mathcal{L}$ is the mean squared error (MSE) between the estimated video $\bm{x}^{K}$ and the ground-truth video $\bm{x}^{\ast}$. 

In the experiments, we use FastDVDnet~\cite{Tassano2020-os} as the denoising method $\mathcal{D}_{\sigma_{k}}(\cdot)$. 
Since there are 60 FastDVDnet layers in the unfolded networks, we require huge GPU memory for the usual training. 
We thus use the checkpointing approach~\cite{Chen2016-mr,Gruslys2016-gz} to reduce GPU memory usage, which can be implemented with a function of PyTorch. 
In regular backpropagation calculations, all the intermediate results from the forward pass are stored for use in the backward pass. 
Checkpointing reduces memory usage by storing only selected intermediate results from the forward pass.
For the non-stored results, recalculation is performed during the backward pass using the stored intermediate results. 
Consequently, this approach requires longer training time than the normal approach.
\section{Simulation Results and Discussion} \label{sec:simulation}
In this section, we evaluate the performance of the PnP-based methods with the proposed noise level parameter tuning via computer simulations. 
Although there are various methods for the SCI reconstruction, the aim of the simulation here is to demonstrate the effect of the parameter settings. 
Therefore, we do not focus on a performance comparison with other SCI reconstruction methods. 
Note again that the proposed approach can be combined with some other approaches, as discussed in Section~\ref{subsec:related}. 
The simulation is conducted by using a computer with Intel Core i9-13900, 64GB memory, and NVIDIA RTX A4500. 
The simulation code is implemented in PyTorch 2.2. 
\subsection{Trained Noise Level Parameters and Average Reconstruction Accuracy}
\begin{figure}[!t]
    \centering
    \includegraphics[width=1\hsize]{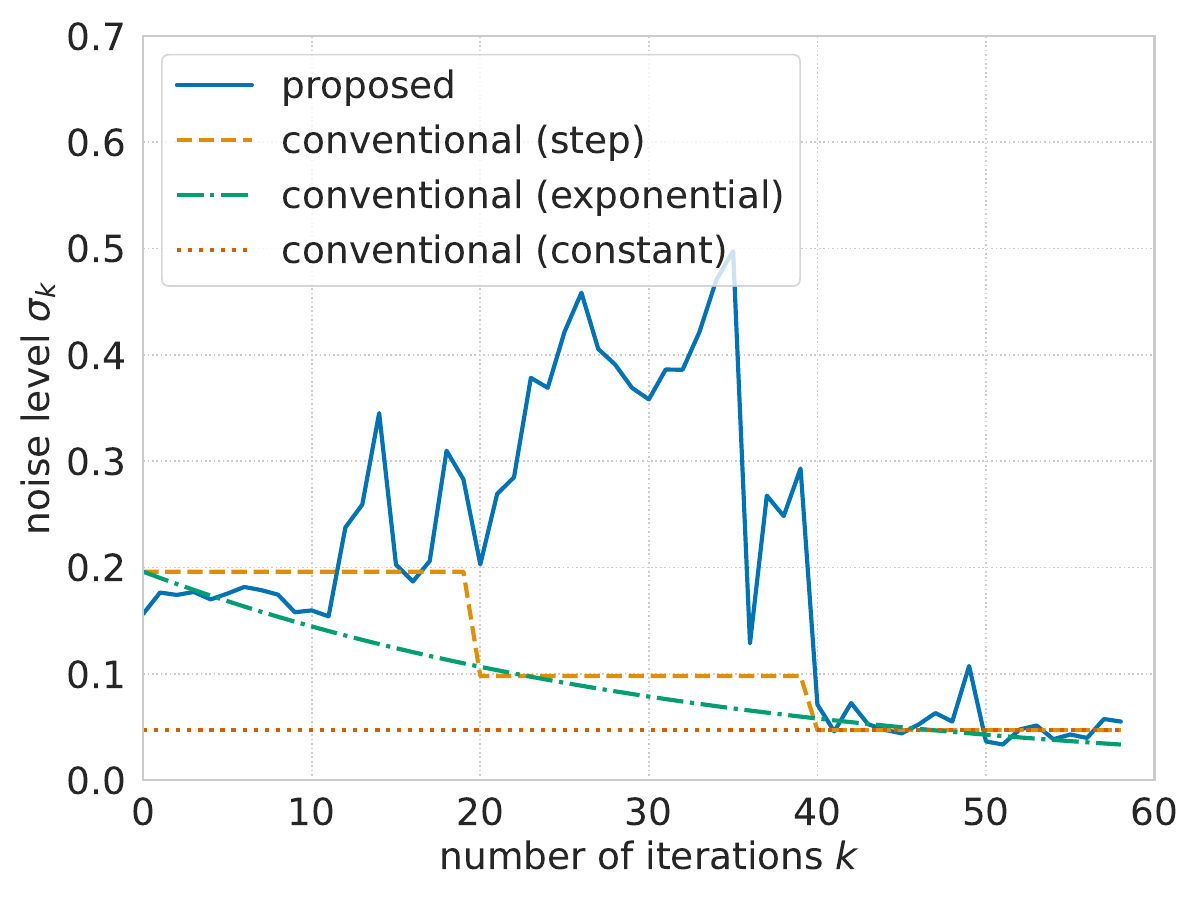}
    \caption{Noise level parameters in PnP-ADMM.}
    \label{fig:noise-level_PnP-ADMM}
\end{figure}

We first evaluate the performance of the proposed parameter tuning for PnP-ADMM. 
Figure~\ref{fig:noise-level_PnP-ADMM} shows the noise level parameters $\curbra{\sigma_{k}}_{k=0,\dotsc,K-1}$ obtained by the proposed method via deep unfolding. 
In the figure, `proposed' represents the noise level parameters obtained by the proposed method. 
For comparison, we also show the noise level parameters in the first, second, and third setting in Section~\ref{subsec:param} as `conventional (step)', `conventional (exponential)', and `conventional (constant)', respectively. 
Note that the noise level parameters in the first setting are the initial values in the training. 
In the simulations, we use the mask $\bm{M}$ in the dataset of the six grayscale videos. 
From Fig.~\ref{fig:noise-level_PnP-ADMM}, we can see that the noise level parameters change significantly by the training. 

The trained parameters is not necessarily monotonically decreasing, though the convergence of PnP-ADMM is usually guaranteed for a decreasing sequence of noise level parameters~\cite{Chan2017-xv,Yuan2022-nj,Shi2023-jw,Shi2024-uz}. 
As shown in later, however, the proposed noise level parameters can achieve a higher reconstruction accuracy than the conventional ones. 
This result implies that monotonically decreasing noise level parameters might not necessarily be optimal in terms of the final reconstruction accuracy. 
A phenomenon similar to the zigzag behavior of the parameters has also been observed in the training of the TISTA~\cite{Ito2019-fm}, which is a deep unfolding-based method for sparse signal recovery. 
Moreover, in reinforcement learning-based parameter tuning for PnP methods~\cite{Wei2020-is}, the resultant noise level parameters are not necessarily monotonically decreasing for medical resonance imaging (MRI). 

\begin{figure}[!t]
    \centering
    \includegraphics[width=1\hsize]{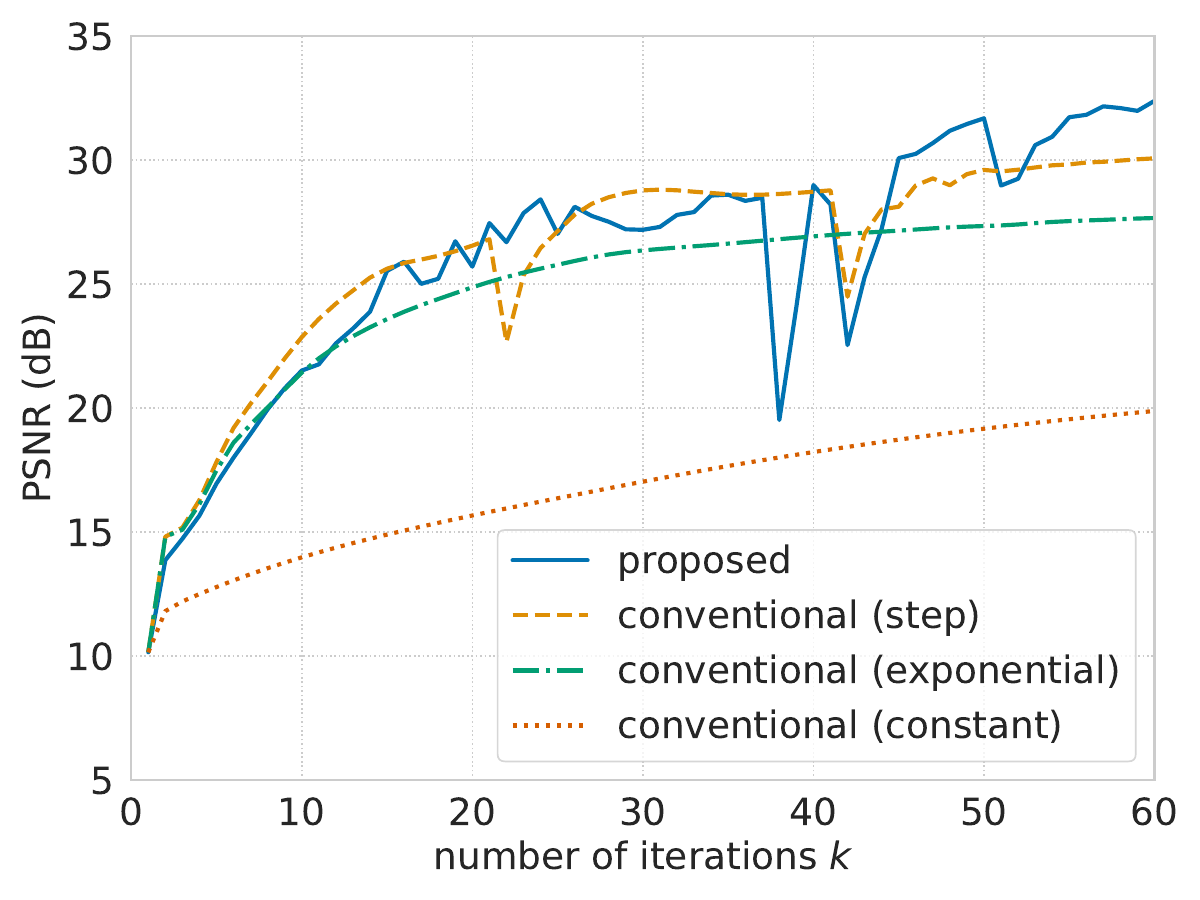}
    \caption{PSNR of PnP-ADMM with various parameter settings.}
    \label{fig:PnP-ADMM}
\end{figure}

Figure~\ref{fig:PnP-ADMM} shows the reconstruction accuracy of PnP-ADMM for various parameter settings. 
In the figure, `proposed' represents the peak signal-to-noise ratio (PSNR) of PnP-ADMM with the noise level parameter obtained by the proposed deep unfolding approach. 
On the other hand, `conventional (step)', `conventional (exponential)', and `conventional (constant)' represent the performance of PnP-ADMM with the noise level parameters in the first, second, and third setting in Section~\ref{subsec:param}, respectively. 
The PSNR is calculated by averaging the result over the test data composed of six videos. 
For all videos, $N_{t} = 8$ frames of the video are reconstructed from a single measurement image. 
The standard deviation of the Gaussian measurement noise is $0.01$, which is the same in the training. 
From Fig.~\ref{fig:PnP-ADMM}, we observe that the noise level parameters significantly affect the reconstruction accuracy. 
We can also see that the proposed parameter tuning can achieve higher reconstruction accuracy than the conventional ones, though the PSNR is not monotonically increasing because of the zig-zag behavior of the noise level parameters. 

\begin{figure}[!t]
    \centering
    \includegraphics[width=1\hsize]{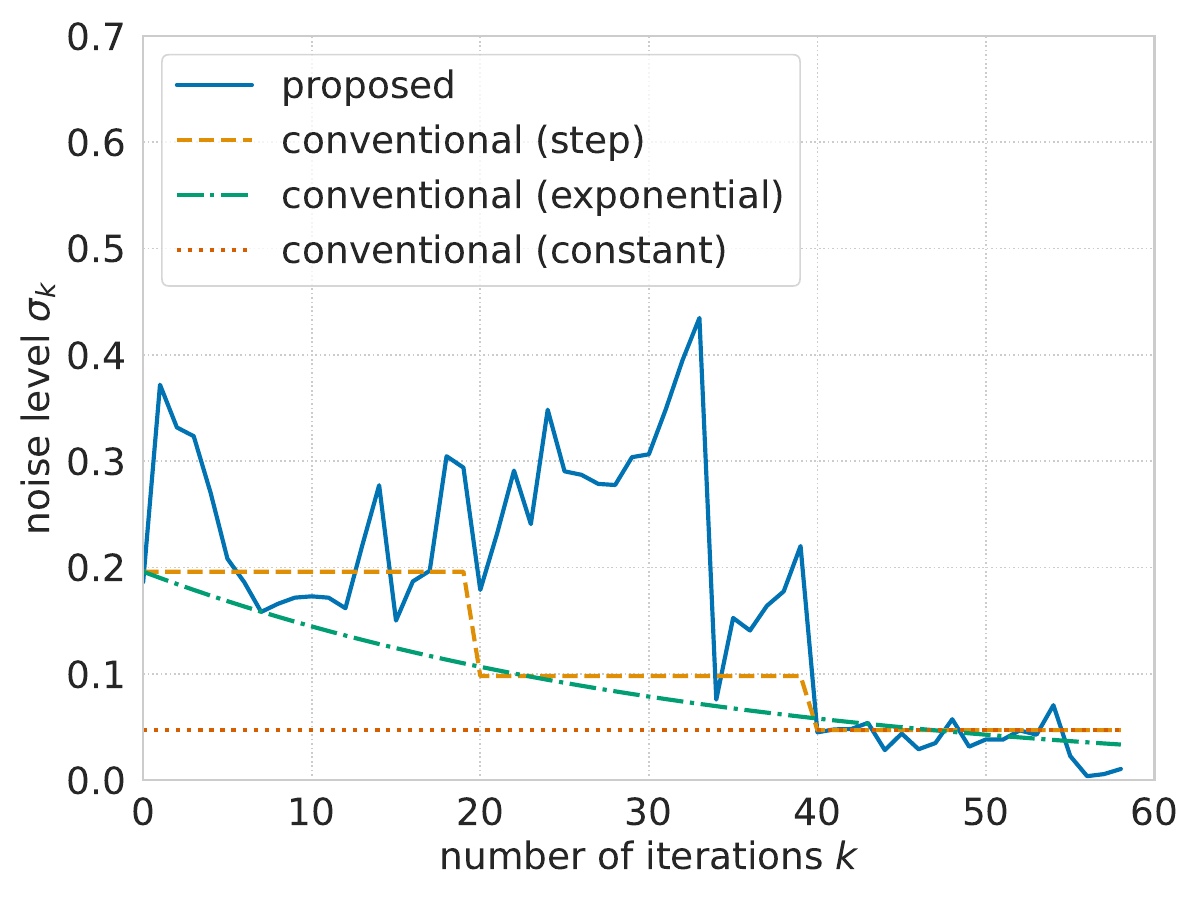}
    \caption{Noise level parameters in PnP-GAP.}
    \label{fig:noise-level_PnP-GAP}
\end{figure}

Next, we evaluate the effectiveness of the proposed approach for PnP-GAP. 
Figure~\ref{fig:noise-level_PnP-GAP} shows the noise level parameters $\curbra{\sigma_{k}}_{k=0,\dotsc,K-1}$ obtained by the proposed method via deep unfolding for PnP-GAP. 
\begin{figure}[!t]
    \centering
    \includegraphics[width=1\hsize]{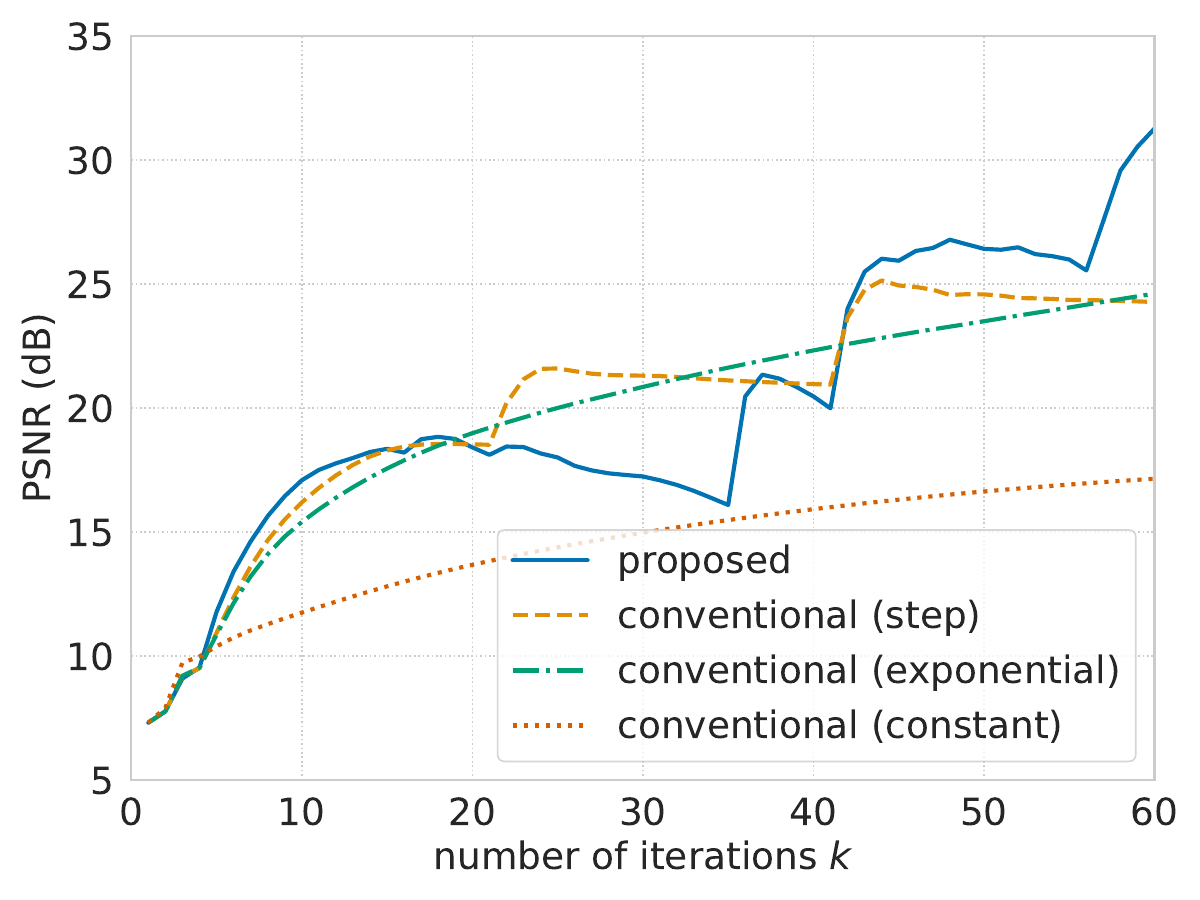}
    \caption{PSNR of PnP-GAP with various parameter settings.}
    \label{fig:PnP-GAP}
\end{figure}
As in the case of PnP-ADMM, the noise level parameters change significantly by the training. 
The resultant noise level parameters are not necessarily monotonically decreasing and a zig-zag behavior is observed. 
This result suggests that the non-decreasing noise level parameters are better in terms of the final reconstruction accuracy also in the case of PnP-GAP. 

Figure~\ref{fig:PnP-GAP} shows the reconstruction accuracy of PnP-GAP for various parameter settings.
As in the case with PnP-ADMM, the PSNR is calculated by averaging the results over the test data of six grayscale videos. 
The standard deviation of the Gaussian measurement noise is $0.01$. 
Figure~\ref{fig:PnP-GAP} shows that the proposed parameter tuning can also achieve a higher reconstruction accuracy than the conventional ones for PnP-GAP, though the PSNR is not increasing monotonically. 
\subsection{Reconstruction Results for Each Video}
\begin{figure*}[p]
    \centering
    \includegraphics[width=1\hsize]{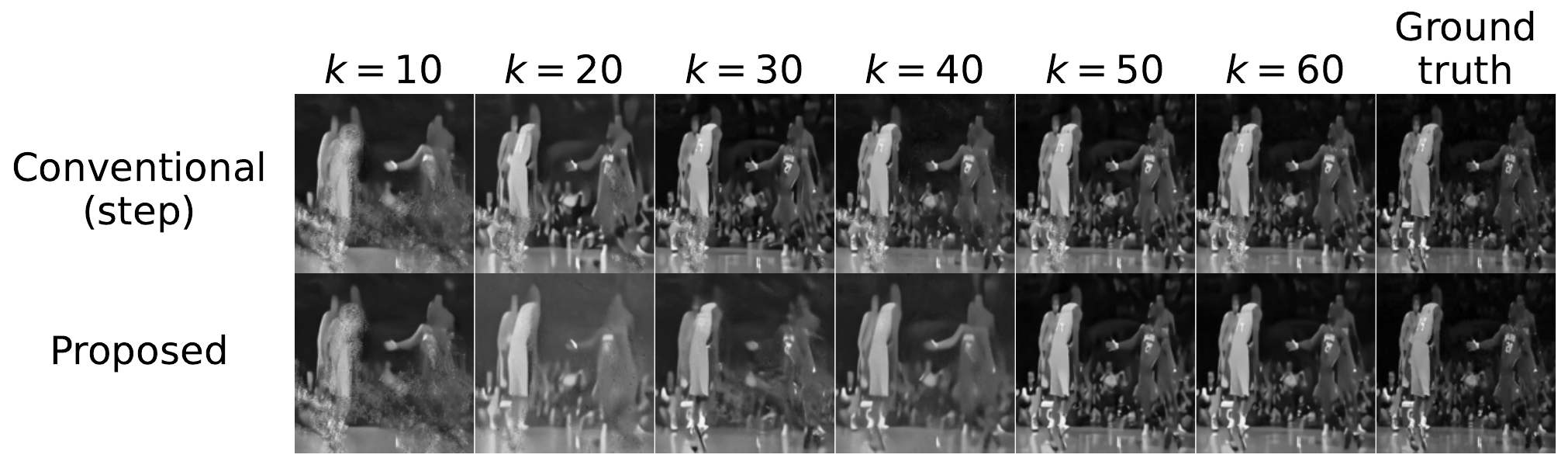}
    \caption{Reconstruction progress of PnP-ADMM per iteration for a frame of \texttt{kobe}. \newline (PSNR (dB)/SSIM of conventional (step): 25.81/0.884, PSNR (dB)/SSIM of proposed: 29.89/0.908)}
    \label{fig:kobe32_ADMM}
\end{figure*}

\begin{figure*}[p]
    \centering
    \includegraphics[width=1\hsize]{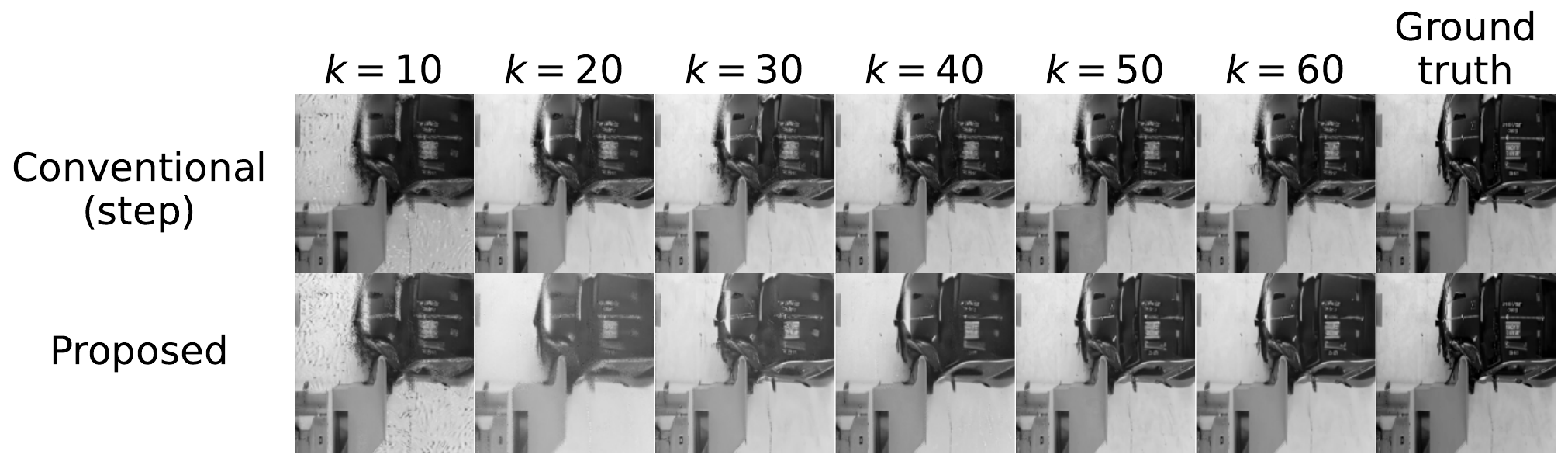}
    \caption{Reconstruction progress of PnP-ADMM per iteration for a frame of \texttt{crash}. \newline (PSNR (dB)/SSIM of conventional (step): 23.39/0.857, PSNR (dB)/SSIM of proposed: 26.78/0.905)}
    \label{fig:crash32_ADMM}
\end{figure*}

In Figs.~\ref{fig:kobe32_ADMM} and~\ref{fig:crash32_ADMM}, we show the reconstruction progress of PnP-ADMM per iteration for a frame of \texttt{kobe} and \texttt{crash}, respectively. 
The standard deviation of the Gaussian measurement noise is $0.01$. 
In the figures, the top row shows the estimate at the $k$-th iteration ($k=10, 20, \dotsc, 60$) with the conventional first parameter setting (step) and the corresponding ground truth image. 
The bottom row shows the estimate obtained using the proposed noise level parameters. 
From the figures, the reconstructed result of PnP-ADMM improves as the iteration proceeds. 
We also observe that the proposed noise level parameters can reconstruct the details of the scene better than the conventional ones. 
For the frame of \texttt{kobe}, the PSNR and structural similarity index measure (SSIM) at $k=60$ for the conventional parameters are 25.81 dB and 0.884, respectively, whereas they are 29.89 dB and 0.908 for the proposed parameters. 
For the frame of \texttt{crash}, the PSNR and SSIM at $k=60$ for the conventional parameters are 23.39 dB and 0.857, respectively, whereas 26.78 dB and 0.905 for the proposed parameters. 
From the quantitative evaluation, we can also see that the proposed noise level parameters can achieve better reconstruction accuracy than the conventional ones. 

\begin{figure*}[p]
    \centering
    \includegraphics[width=1\hsize]{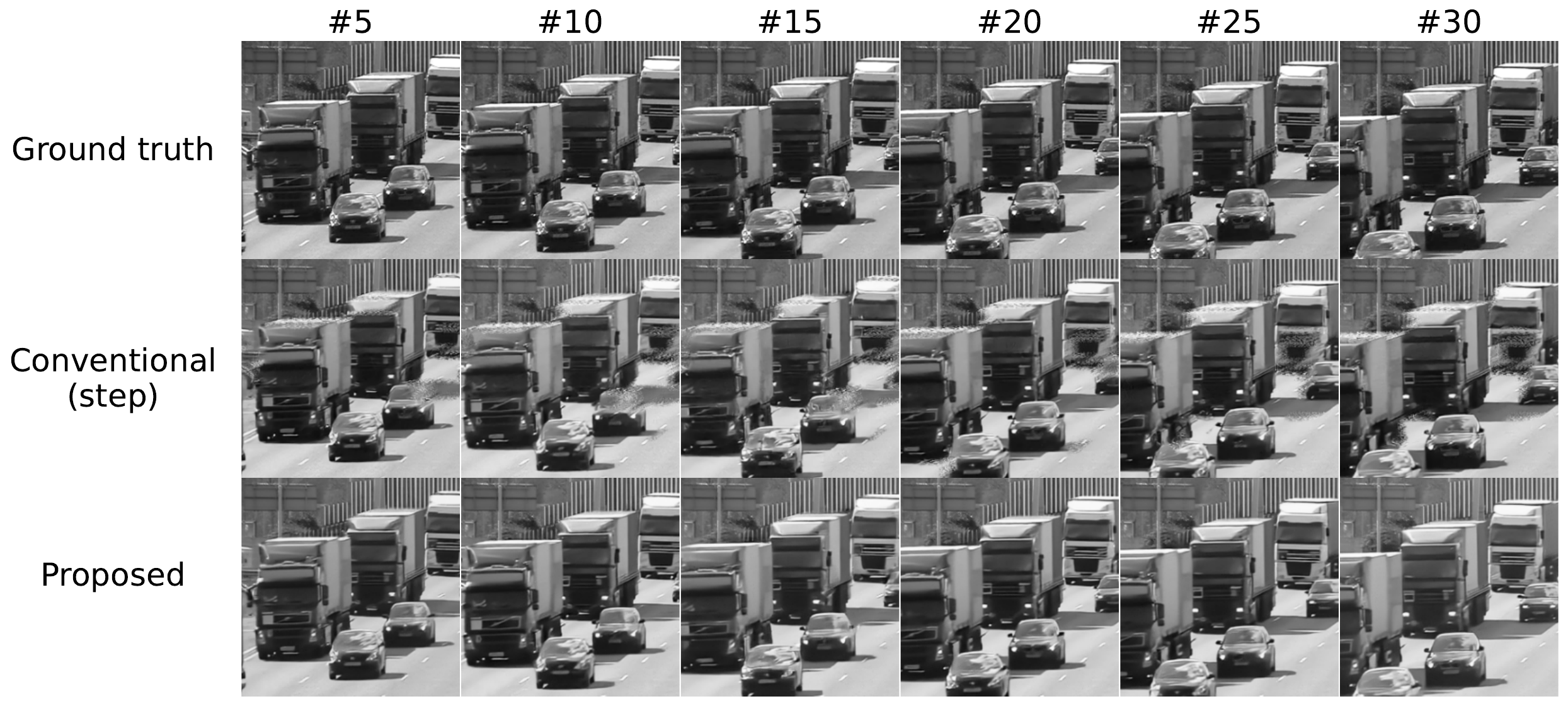}
    \caption{Reconstruction results of PnP-ADMM for \texttt{traffic}.}
    \label{fig:traffic48_ADMM}
\end{figure*}

Figure~\ref{fig:traffic48_ADMM} shows an example of the reconstruction results of PnP-ADMM for \texttt{traffic}. 
The standard deviation of the Gaussian measurement noise is $0.01$, which is the same as that in the training. 
In the figure, the top row shows the ground-truth, the middle row shows the reconstruction results of the conventional first parameter setting (step) in Section~\ref{subsec:param}, and the bottom row shows the reconstruction results of the proposed parameters. 
The numbers above the images represent the indices of the frames. 
We can see that the proposed parameter tuning can achieve better reconstruction results than conventional ones. 
In particular, although the conventional parameters fail to reconstruct the details of the cars, the proposed parameters can reconstruct the details more clearly. 

A possible reason for the performance degradation of the conventional parameters is that the initial value of the algorithm is simply set as $\bm{v}^{0} = \bm{\Phi}^{\top} \bm{y}$ for the simple comparison. 
If we use the initialization with GAP-total variation (GAP-TV) as in~\cite{Yuan2022-nj}, the performance of the conventional parameters would be improved. 
By the proposed approach, however, we can obtain good reconstruction results without such initialization and reduce the computational cost. 
It should be noted that the proposed parameter tuning can also be combined with such initialization method. 

TABLE~\ref{tab:PSNR_SSIM_ADMM_noisy0.01} shows the PSNR and SSIM of PnP-ADMM for \texttt{kobe}, \texttt{traffic}, \texttt{runner}, \texttt{drop}, \texttt{crash}, and \texttt{aerial}. 
\begin{table*}[!t]
    \centering
    \caption{PSNR (left entry) and SSIM (right entry) of PnP-ADMM. (noisy case, standard deviation: $0.01$)}
    \label{tab:PSNR_SSIM_ADMM_noisy0.01}
    \begin{tabular}{l|cccccc|c}
    parameter setting & \texttt{kobe} & \texttt{traffic} & \texttt{runner} & \texttt{drop} & \texttt{crash} & \texttt{aerial} & average \\ \hline
    conventional (step) & 29.83/0.924 & 22.73/0.857 & 34.24/0.946 & 41.02/0.981 & 25.16/0.894 & 27.40/0.880 & 30.06/0.914 \\
    conventional (exponential) & 25.11/0.853 & 19.69/0.731 & 31.71/0.920 & 39.24/0.971 & 24.36/0.867 & 25.83/0.849 & 27.66/0.865 \\
    conventional (constant) & 22.40/0.790 & 17.60/0.570 & 24.26/0.849 & 16.90/0.283 & 17.63/0.474 & 20.50/0.693 & 19.88/0.610 \\
    proposed & \textbf{32.32}/\textbf{0.938} & \textbf{28.43}/\textbf{0.930} & \textbf{35.45}/\textbf{0.949} & \textbf{41.51}/\textbf{0.986} & \textbf{28.04}/\textbf{0.924} & \textbf{28.54}/\textbf{0.887} & \textbf{32.38}/\textbf{0.936} \\
    \end{tabular}
\end{table*}
TABLE~\ref{tab:PSNR_SSIM_GAP_noisy0.01} lists the values of PnP-GAP. 
\begin{table*}[!t]
    \centering
    \caption{PSNR (left entry) and SSIM (right entry) of PnP-GAP. (noisy case, standard deviation: $0.01$)}
    \label{tab:PSNR_SSIM_GAP_noisy0.01}
    \begin{tabular}{l|cccccc|c}
    parameter setting & \texttt{kobe} & \texttt{traffic} & \texttt{runner} & \texttt{drop} & \texttt{crash} & \texttt{aerial} & average \\ \hline
    conventional (step) & 25.88/0.608 & 21.03/0.619 & 25.75/0.511 & 26.20/0.472 & 23.00/0.513 & 23.77/0.491 & 24.27/0.536 \\
    conventional (exponential) & 25.29/0.647 & 20.20/0.614 & 27.03/0.572 & 27.58/0.538 & 23.35/0.557 & 24.22/0.540 & 24.61/0.578 \\
    conventional (constant) & 19.42/0.457 & 15.37/0.365 & 20.40/0.444 & 15.69/0.182 & 15.82/0.282 & 16.22/0.353 & 17.15/0.347 \\
    proposed & \textbf{31.90}/\textbf{0.914} & \textbf{28.01}/\textbf{0.912} & \textbf{34.36}/\textbf{0.922} & \textbf{37.72}/\textbf{0.921} & \textbf{27.75}/\textbf{0.890} & \textbf{27.84}/\textbf{0.831} & \textbf{31.26}/\textbf{0.898} \\
    \end{tabular}
\end{table*}
The standard deviation of the Gaussian noise is $0.01$. 
From the tables, we can see that the proposed parameter tuning can achieve a higher reconstruction accuracy than the conventional parameters for each video. 

Finally, we evaluate the robustness of the proposed parameter against the standard deviation of the measurement noise. 
TABLE~\ref{tab:PSNR_SSIM_ADMM_noiseless} shows the PSNR and SSIM of PnP-ADMM when the standard deviation of the Gaussian measurement noise is $0$, that is, $\bm{e}=\bm{0}$. 
\begin{table*}[!t]
    \centering
    \caption{PSNR (left entry) and SSIM (right entry) of PnP-ADMM. (noiseless case)}
    \label{tab:PSNR_SSIM_ADMM_noiseless}
    \begin{tabular}{l|cccccc|c}
    parameter setting & \texttt{kobe} & \texttt{traffic} & \texttt{runner} & \texttt{drop} & \texttt{crash} & \texttt{aerial} & average \\ \hline
    conventional (step) & 29.99/0.929 & 23.01/0.867 & 34.53/\textbf{0.950} & \textbf{41.90}/\textbf{0.989} & 25.26/0.900 & 27.41/0.875 & 30.35/0.918 \\
    conventional (exponential) & 25.35/0.859 & 19.75/0.742 & 31.91/0.924 & 39.35/0.976 & 24.45/0.873 & 25.91/0.857 & 27.79/0.872 \\
    conventional (constant) & 22.41/0.796 & 17.59/0.572 & 24.06/0.847 & 16.89/0.283 & 17.63/0.477 & 20.50/0.697 & 19.85/0.612 \\
    proposed & \textbf{32.40}/\textbf{0.939} & \textbf{28.47}/\textbf{0.931} & \textbf{35.49}/0.949 & 41.34/0.983 & \textbf{28.07}/\textbf{0.924} & \textbf{28.50}/\textbf{0.878} & \textbf{32.38}/\textbf{0.934} \\
    \end{tabular}
\end{table*}
TABLE~\ref{tab:PSNR_SSIM_GAP_noiseless} shows those of PnP-GAP. 
\begin{table*}[!t]
    \centering
    \caption{PSNR (left entry) and SSIM (right entry) of PnP-GAP. (noiseless case)}
    \label{tab:PSNR_SSIM_GAP_noiseless}
    \begin{tabular}{l|cccccc|c}
    parameter setting & \texttt{kobe} & \texttt{traffic} & \texttt{runner} & \texttt{drop} & \texttt{crash} & \texttt{aerial} & average \\ \hline
    conventional (step) & 26.37/0.647 & 21.26/0.642 & 26.88/0.575 & 27.94/0.564 & 23.40/0.558 & 24.01/0.524 & 24.97/0.585 \\
    conventional (exponential) & 25.62/0.676 & 20.33/0.632 & 27.72/0.616 & 28.02/0.571 & 23.56/0.590 & 24.47/0.570 & 24.95/0.609 \\
    conventional (constant) & 19.49/0.490 & 15.36/0.370 & 20.43/0.488 & 15.69/0.187 & 15.83/0.293 & 16.25/0.379 & 17.18/0.368 \\
    proposed & \textbf{32.08}/\textbf{0.920} & \textbf{28.08}/\textbf{0.915} & \textbf{34.34}/\textbf{0.918} & \textbf{38.01}/\textbf{0.926} & \textbf{27.66}/\textbf{0.878} & \textbf{28.13}/\textbf{0.835} & \textbf{31.38}/\textbf{0.899} \\
    \end{tabular}
\end{table*}
Moreover, TABLEs~\ref{tab:PSNR_SSIM_ADMM_noisy0.05} and~\ref{tab:PSNR_SSIM_GAP_noisy0.05} show the performance of PnP-ADMM and PnP-GAP when the standard deviation of the Gaussian measurement noise is $0.05$, respectively.
\begin{table*}[!t]
    \centering
    \caption{PSNR (left entry) and SSIM (right entry) of PnP-ADMM. (noisy case, standard deviation: $0.05$)}
    \label{tab:PSNR_SSIM_ADMM_noisy0.05}
    \begin{tabular}{l|cccccc|c}
    parameter setting & \texttt{kobe} & \texttt{traffic} & \texttt{runner} & \texttt{drop} & \texttt{crash} & \texttt{aerial} & average \\ \hline
    conventional (step) & 26.17/0.785 & 20.57/0.729 & 30.07/0.798 & 32.89/0.817 & 23.86/0.764 & 26.18/0.753 & 26.62/0.774 \\
    conventional (exponential) & 22.14/0.644 & 18.74/0.600 & 28.25/0.779 & 31.76/0.809 & 23.48/0.739 & 24.68/0.717 & 24.84/0.715 \\
    conventional (constant) & 21.88/0.671 & 17.46/0.515 & 24.11/0.734 & 16.88/0.272 & 17.50/0.419 & 20.46/0.621 & 19.72/0.539 \\
    proposed & \textbf{30.22}/\textbf{0.873} & \textbf{27.50}/\textbf{0.893} & \textbf{32.91}/\textbf{0.889} & \textbf{36.42}/\textbf{0.918} & \textbf{27.45}/\textbf{0.868} & \textbf{27.93}/\textbf{0.841} & \textbf{30.41}/\textbf{0.880} \\
    \end{tabular}
\end{table*}
\begin{table*}[!t]
    \centering
    \caption{PSNR (left entry) and SSIM (right entry) of PnP-GAP. (noisy case, standard deviation: $0.05$)}
    \label{tab:PSNR_SSIM_GAP_noisy0.05}
    \begin{tabular}{l|cccccc|c}
    parameter setting & \texttt{kobe} & \texttt{traffic} & \texttt{runner} & \texttt{drop} & \texttt{crash} & \texttt{aerial} & average \\ \hline
    conventional (step) & 23.67/0.509 & 19.49/0.524 & 23.84/0.414 & 24.73/0.408 & 21.83/0.423 & 22.78/0.426 & 22.72/0.451 \\
    conventional (exponential) & 22.64/0.512 & 18.99/0.512 & 25.26/0.502 & 26.32/0.488 & 22.56/0.485 & 23.17/0.468 & 23.16/0.494 \\
    conventional (constant) & 19.10/0.390 & 15.27/0.333 & 20.08/0.351 & 15.60/0.169 & 15.70/0.253 & 16.08/0.305 & 16.97/0.300 \\
    proposed & \textbf{29.36}/\textbf{0.803} & \textbf{26.84}/\textbf{0.849} & \textbf{31.23}/\textbf{0.803} & \textbf{33.52}/\textbf{0.816} & \textbf{26.87}/\textbf{0.785} & \textbf{27.37}/\textbf{0.757} & \textbf{29.20}/\textbf{0.802} \\
    \end{tabular}
\end{table*}
From the tables, we can see that the proposed parameter tuning can achieve higher reconstruction accuracy than the conventional ones for most videos, even though the parameter is trained by using the data with the standard deviation of $0.01$. 
This result suggests that the proposed parameter tuning is robust to changes in the standard deviation of the measurement noise. 
Although the performance in the noiseless case is slightly worse than that in the noisy case with $\sigma=0.01$ for some videos, this would be because the noise level parameters are trained by using the noisy data. 
\subsection{Evaluation for Real Data}
\begin{figure*}[t]
    \centering
    \includegraphics[width=1\hsize]{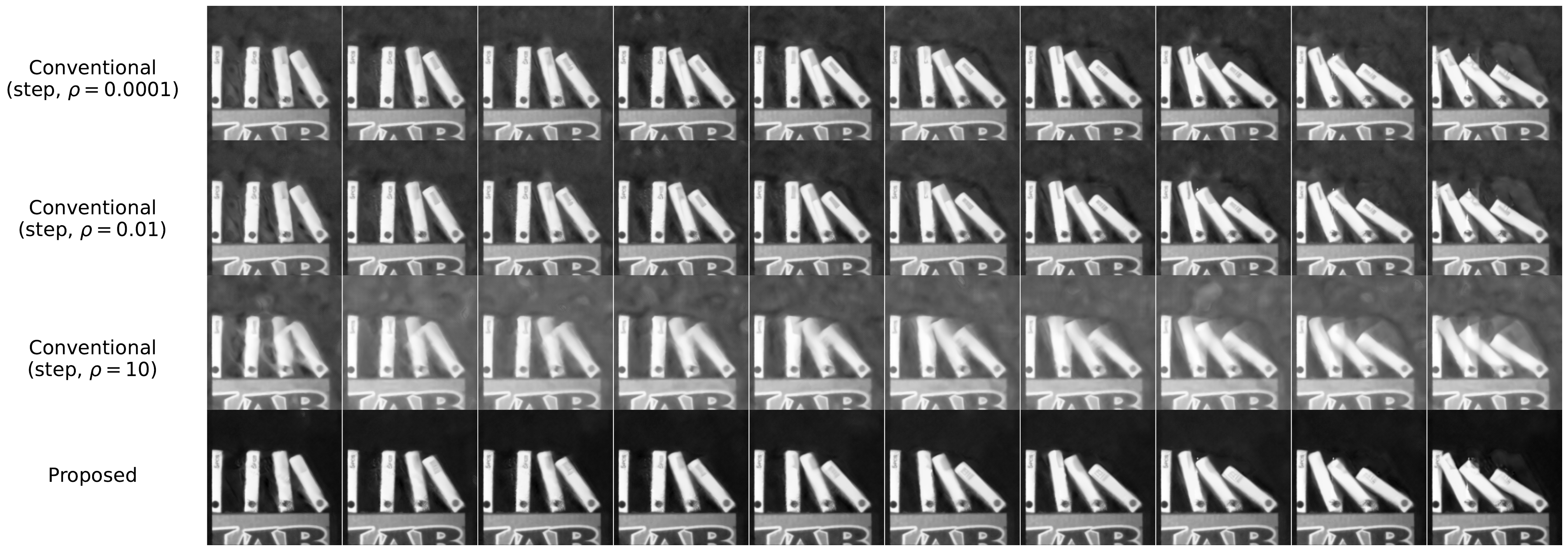}
    \caption{Reconstruction results of PnP-ADMM for real data \texttt{domino} ($N_{t} = 10$).}
    \label{fig:real_data_domino}
\end{figure*}

\begin{figure*}[t]
    \centering
    \includegraphics[width=1\hsize]{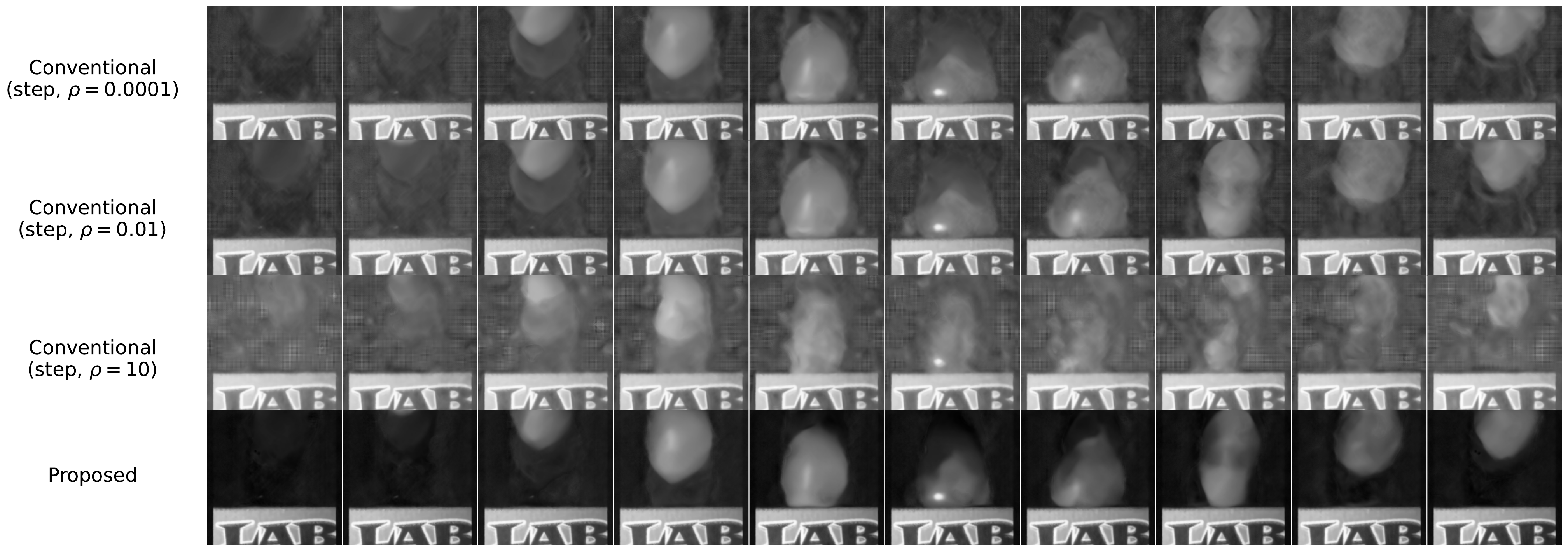}
    \caption{Reconstruction results of PnP-ADMM for real data \texttt{water balloon}  ($N_{t} = 10$).}
    \label{fig:real_data_waterBalloon}
\end{figure*}

Finally, we evaluate the performance of the proposed parameter tuning method for real data~\cite{Qiao2020-xe}. 
Figures~\ref{fig:real_data_domino} and~\ref{fig:real_data_waterBalloon} show the reconstruction results of PnP-ADMM for a real video \texttt{domino} and \texttt{water balloon}, respectively. 
In the figures, the top three rows show the reconstruction results for each frame of the conventional first parameter setting (step) in Section~\ref{subsec:param}, and the bottom row shows the reconstruction results of the proposed parameters. 
For the conventional ones, we show the results for $\rho=0.0001, 0.01$, and $10$. 
For the proposed method, we show only the results with $\rho=0.01$ because the training is performed with $\rho=0.01$ as described in Section~\ref{subsubsec:training}. 
In the experiments, $N_{t} = 10$ frames of the video are reconstructed from a single compressed image. 
To obtain better reconstruction results, we use the initialization with GAP-TV for the real data. 
From Figs.~\ref{fig:real_data_domino} and~\ref{fig:real_data_waterBalloon}, we can see that the proposed parameter tuning can achieve better reconstruction results than the conventional ones for real data, even though the parameter is trained by using the synthetic data. 
However, since there are still some artifacts in the reconstruction results (e.g., the latter frames in the \texttt{domino} video), further improvement would be required for the real data. 
One possible approach is to use the joint denoising strategy~\cite{Qiao2020-xe}, where the weighted sum of multiple denoisers is used in the denoinsing process. 
Although we do not focus on the joint denoising strategy in this paper, the proposed parameter tuning can also be combined with the strategy.
\section{Conclusion} \label{sec:conclusion}
In this study, we have addressed the critical challenge of designing noise level parameters $\curbra{\sigma_{k}}_{k=0,\dotsc,K-1}$ in PnP-based video SCI algorithms. 
To this end, we have proposed a data-driven tuning approach leveraging the deep unfolding framework, which systematically learns the parameters by minimizing a loss function derived from the reconstruction task. 
Through extensive simulations, we demonstrated that the proposed method outperforms the conventional manually designed parameters in the reconstruction accuracy. 
Remarkably, the trained parameters maintained robust performance even under varying levels of measurement noise, highlighting their adaptability and effectiveness. 
Since the performance improvement is obtained by the choice of the noise level parameters only, our results demonstrate the importance of the parameter tuning in PnP-based video SCI. 

An intriguing finding of this study is the non-monotonic behavior of the learned noise level parameters, which diverges from the conventional assumption of monotonicity required for several convergence analyses in PnP-based methods. 
This result implies that higher reconstruction accuracy can be achieved with parameter settings that deviate from these assumptions. 
To bridge this gap, future work should focus on developing theoretical frameworks to explain the effectiveness of non-monotonic parameter sequences and exploring adaptive tuning strategies that dynamically optimize parameters during the iterative process. 
Such efforts could enhance the robustness and applicability of the PnP methods. 

Future work also includes the training of other parameters such as $\rho$ in PnP-ADMM, examination of the robustness to the mask, and extension for the reconstruction of color videos. 
The integration of the proposed parameter tuning with other SCI reconstruction methods is also an interesting topic for further development of SCI. 
Addressing these challenges would improve the generality and applicability of the proposed method in practical scenarios. 
%
%------------------
% References
%------------------
\bibliographystyle{IEEEtran}
\bibliography{paperpile}
\newpage
\begin{IEEEbiographynophoto}{Takashi Matsuda} 
received the bachelor’s degree and master’s degree in engineering from Osaka University in 2021 and 2023, respectively. 
His research interests include snapshot compressive imaging and deep unfolding. 
\end{IEEEbiographynophoto}

\begin{IEEEbiographynophoto}{Ryo Hayakawa} (Member, IEEE) 
received the bachelor’s degree in engineering, master’s degree in informatics, and Ph.D. degree in informatics from Kyoto University in 2015, 2017, and 2020, respectively. 
He was a Research Fellow (DC1) of the Japan Society for the Promotion of Science (JSPS) from 2017 to 2020, and an Assistant Professor at Osaka Univeristy from 2020 to 2023. 
From 2023, he is an Associate Professor at Institute of Engineering, Tokyo University of Agriculture and Technology. 
From 2023, he is an Associate Editor of IEICE Transactions on Fundamentals of Electronics, Communications and Computer Sciences. 
He received the 33rd Telecom System Technology Student Award, APSIPA ASC 2019 Best Special Session Paper Nomination Award, and the 16th IEEE Kansai Section Student Paper Award. 
His research interests include signal processing and mathematical optimization. 
He is a member of IEEE and IEICE.
\end{IEEEbiographynophoto}

\begin{IEEEbiographynophoto}{Youji Iiguni} (Member, IEEE) 
received the B.E. and M.E. degrees from Kyoto University, Kyoto, Japan, in 1982 and 1984, respectively, and the D.E. degree from Kyoto University in 1989. 
He was an Assistant Professor with Kyoto University from 1984 to 1995, and an Associate Professor with Osaka University, Osaka, Japan. 
Since 2003, he has been a Professor with Osaka University. 
His research interest includes signals and systems analysis.
\end{IEEEbiographynophoto}

\vfill

\end{document}